\documentclass[usenatbib,iop,numberedappendix]{aeb_emulateapj_2010}

\usepackage[usenames,dvipsnames]{color}
\usepackage{hyperref}
\definecolor{darkblue}{rgb}{0.0,0.0,0.3}
\hypersetup{colorlinks,breaklinks,
            linkcolor=darkblue,urlcolor=darkblue,
            anchorcolor=darkblue,citecolor=darkblue}
\DeclareSymbolFont{cmletters}{OML}{cmm}{m}{it}
\DeclareMathSymbol{v}{\mathalpha}{cmletters}{"76}

\usepackage{amsmath}
\usepackage{amssymb}
\usepackage{xspace}
\usepackage[normalem]{ulem}

\usepackage{natbib}

\DeclareMathOperator{\ThetaFunc}{\Theta}
\def\del#1{{}}

\defcitealias{2013MNRAS.429...20T}{TA13}

\def\s{{\rm s}} %...........seconds
\def\Ms{{\rm M}\s} %........megaseconds
\def\yr{{\rm yr}} %.........years
\def\Hz{{\rm Hz}} %.........Hertz
\def\GHz{{\rm GHz}} %.......Gigahertz
\def\THz{{\rm THz}} %.......Terahertz

\def\m{{\rm m}} %...........meters
\def\mm{{\rm m}\m} %........millimeters
\def\cm{{\rm c}\m} %........centimeters
\def\pc{{\rm pc}} %.........parsecs
\def\kpc{{\rm k}\pc} %......kiloparsecs
\def\Mpc{{\rm M}\pc} %......megaparsecs
\def\Ms{M_\odot} %...........solar masses

\def\eV{{\rm eV}} %.........electron volts
\def\meV{{\rm m}\eV} %......kiloelectron volts
\def\GeV{{\rm G}\eV} %......gigaelectron volts
\def\TeV{{\rm T}\eV} %......teraelectron volts
\def\erg{{\rm erg}} %.......ergs
\def\Jy{{\rm Jy}} %.........Janksy (1e-26 J/sHzm^2 = 1e-23 erg/sHzcm^2)
\def\mJy{{\rm m}\Jy} %......milliJansky

\def\G{{\rm G}} %...........gauss
\def\muas{\mu{\rm as}} %....micro-arcseconds
\def\Ms{{M_\odot}}

\def\Rg{{r_{\rm g}}}

\newcommand\bmath[1] {\mbox{\boldmath$\rm #1$}}

\begin{document}

\title{
Horizon-scale lepton acceleration in jets: Explaining the compact
radio emission in M87
}

\author{
Avery E.~Broderick,\altaffilmark{1,2}
and
Alexander Tchekhovskoy\,\altaffilmark{3,4}
}
\altaffiltext{1}{Perimeter Institute for Theoretical Physics, 31 Caroline Street North, Waterloo, ON, N2L 2Y5, Canada}
\altaffiltext{2}{Department of Physics and Astronomy, University of Waterloo, 200 University Avenue West, Waterloo, ON, N2L 3G1, Canada}
\altaffiltext{3}{Departments of Astronomy and Physics, University of
  California Berkeley, Berkeley,  CA 94720-3411, USA; Einstein Fellow}
\altaffiltext{4}{Lawrence Berkeley National Laboratory, 1 Cyclotron Rd,
  Berkeley, CA 94720, USA}

\shorttitle{Horizon-scale lepton acceleration in M87}
\shortauthors{Broderick \& Tchekhovskoy}

\begin{abstract}
It has now become clear that the radio jet in the giant elliptical
galaxy M87 must turn on very
close to the black hole.  This implies the efficient acceleration of
leptons within the jet at scales much smaller than feasible by the
typical dissipative events usually invoked to explain jet synchrotron
emission.  Here we show that the stagnation surface, the separatrix
between material that falls back into the black hole and material that
is accelerated outward forming the jet, is a natural site of pair
formation and particle acceleration.  This occurs via an
inverse-Compton pair catastrophe driven by unscreened electric fields
within the charge-starved region about the stagnation surface and
substantially amplified by a post-gap cascade.  For typical estimates
of the jet properties in M87, we find excellent quantitive agreement
between the predicted relativistic lepton densities and those required
by recent high-frequency radio observations of M87.  This mechanism
fails to adequately fill a putative jet from Sagittarius~A* with
relativistic leptons, which may explain the lack of an obvious
radio jet in the Galactic center.  Finally, this process implies a
relationship between the kinetic jet power and the gamma-ray
luminosity of blazars, produced during the post-gap cascade.
\end{abstract}


\maketitle

\section{Introduction} \label{sec:I}
Roughly 10\% of active galactic nuclei (AGN), exhibit powerful radio
jets.  These can extend for intergalactic distances, and potentially
have impacts for the formation and evolution of their galactic
environments.  Great strides have been made in the past two decades in
the theoretical modeling of the structure and formation of AGN jets,
with current general relativistic magnetohydrodynamic (GRMHD) models now
capable of reproducing the structure and Lorentz factors ($\Gamma$) of
observed objects.  In these, large-scale electromagnetic fields sourced
and collimated by the surrounding accretion flow efficiently extract
the rotational energy of the black hole \citep{1977MNRAS.179..433B,2010ApJ...711...50T,2011MNRAS.418L..79T,2012JPhCS.372a2040T,2012MNRAS.423L..55T,2012MNRAS.423.3083M,2015ASSL..414...45T}. 
Within the context of GRMHD simulations, a canonical outflow structure
has emerged, consisting of a central relativistic jet core surrounded
by a moderately relativistic jet sheath, followed by a
non-relativistic disk wind.

Less certain is why jets shine.  This has become especially pressing
now that it has become clear that the jet in M87 exhibits
horizon-scale structure, implying emission on horizon scales \citep{2012Sci...338..355D}.
It is widely believed that from the radio to the UV, the observed jet
emission is due to synchrotron emission from a population of nonthermal
leptons.  Not presently known is the mechanism by which this
population is produced.  Uncertain are the processes by which jets are
mass loaded, particles within the jet are accelerated, and the
dynamical structures associated with the observed jet at different
wavelengths (e.g., the jet spine vs. jet sheath).  Any candidate
mechanisms must simultaneously explain two critical features.

First, the leptons must be located within the fast region of the
jet.  Radio observations far from the radio core show super-luminal
motions which imply large values of the bulk Lorentz factor, $\Gamma\gg1$.  Thus the synchrotron emitting
particles in particular must be moving at these extreme velocities.
This suggests that the radio emission is not due to a mass-loaded,
trans-relativistic disk wind surrounding and potentially collimating
the jet.  This is problematic because jet formation simulations, as
well as approximate analytical models, imply the presence of
large-scale, dynamically dominant, ordered magnetic fields.
Correspondingly, the exchange of particles onto the jet magnetic field
lines is dramatically suppressed, with the consequence that this is
unlikely to be the mechanism by which jets are mass loaded.
Funnel regions can potentially be filled with pairs via collisions of
gamma-rays formed in the disk.  These gamma-rays are presumably
produced through a sequence of inverse Compton scatterings of disk
sub-mm photons with the hot disk electrons
\citep{2011ApJ...730..123L,2011ApJ...735....9M}, though such a
mechanism requires order unity radiative efficiency of the accretion
flow.  However, for the radiatively inefficient accretion flows
believed to be relevant at the low Eddington ratio observed in objects
like M87, most of the sub-mm emission is produced in the jet, and this process
is suppressed.
Alternative processes, such as 
neutron diffusion \citep{Levi-Eich:03}, have been suggested as a means to accomplish this,
though fail at low source luminosities (e.g., M87). 

\begin{table*}
\begin{center}
\caption{Key Symbol Definitions}
\begin{minipage}{\textwidth}
{\normalsize
\begin{center}
\begin{tabular}{r@{$\quad$}c@{$\quad$}l}
\hline
\hline
Name                & Eq. &  Definition                        \\
\hline
$u_s$               & \ref{eq:us} & Density of seed photons            \\
$n_g$              & \ref{eq:ngap} & Number density of charges in the gap \\
$n_\infty$          & \ref{eq:enhancement} & Number density of charges
                                             at the end of the cascade \\
$\ell_{\rm IC}$      & \ref{eq:ellIC} & Cooling length of electrons due to
                      inverse Compton scattering off seed photons   \\
$\ell_{\gamma\gamma}$ & \ref{eq:ellgg} & Mean free path of $\gamma$-ray photons due to
                        pair production on seed photons\\
$\ell_{\rm C}$      & \ref{eq:ellc}& Cooling length of electrons due to
                      curvature radiation             \\
$\Delta$          & \ref{eq:gapthickness} & Gap thickness \\
$\gamma_{\rm max,IC}$& \ref{eq:gammamaxIC} &  
                       Maximum electron Lorentz factor due to the
                       acceleration over the distance $\ell_{\rm IC}$ \\
$\gamma_{\rm max,C}$ & \ref{eq:gammamaxc} &  Maximum electron
                                           Lorentz factor due to 
                                           the  acceleration over
                                           the distance $\ell_{\rm C}$  \\
$\gamma_{\rm max}$   & \ref{eq:gammamax} & Maximum electron
                                         Lorentz factor due to 
                                         the  acceleration over
                                         the distance $\Delta$    \\
$\gamma_{\gamma\gamma}$ &\ref{eq:ggg} & Lorentz factor of electrons below which
                          they do not encounter enough seed photons
                          to pair produce              \\
$\gamma_\infty$ & \ref{eq:gammainfty} & Lorentz factor of electrons
                          below which they no longer efficiently
                          Compton cool\\
$\Omega_{\rm F}$    & \ref{eq:omegaf}  & Field line angular rotation frequency    \\
$B$                & \ref{eq:BPest}   & Magnetic field strength at the
                                        stagnation surface, $r\sim10r_g$  \\
\hline
\hline
\label{tab1}
\end{tabular}
\end{center}
}
\end{minipage}
\end{center}
\end{table*}

Second, the leptons must be accelerated.  Simply mass loading a jet is
insufficient; assuming the primary emission mechanism is synchrotron
emission, the leptons must also be accelerated to large internal
Lorentz factors as well.  Far from the black hole this may be
accomplished via internal shocks, magnetic reconnection, damping of
plasma waves, etc.  However, already within $10\Rg$ bright radio emission is
detected in M87, where $\Rg\equiv GM/c^2$ is the black hole's
gravitational radius \citep{2012Sci...338..355D}.  Thus, particle
acceleration must  
also be occurring quite close to the black hole, where these processes
either do not occur or are strongly suppressed.

Within the context of M87, there are numerous spectral and
morphological constraints that should inform any attempt to model the
lepton loading and acceleration in jets from low luminosity AGN
(LLAGN; potentially comprising the majority of the population).  Most
importantly, at a wavelength of 7mm and shorter the luminosity from
the extended jet considerably exceeds that from the core, and thus
importantly the luminosity is {\em not} core dominated. 
This morphology is inconsistent with the
core-dominated morphology produced by models in which
the bulk of the accelerated leptons are produced at the black hole,
e.g., models in which nonthermal energy in electron-positron pairs is
$u_{e^\pm}\propto B^2$
\citep{2009ApJ...706..497M,2012ApJ...755..133S,2012MNRAS.421.1517D,2013A&A...559L...3M} 
or with the annihilation of high-energy photons produced in
low-luminosity radiatively-inefficient accretion flows
\citep{2011ApJ...735....9M}.  It is, however, consistent with models
in which the energetic lepton population is generated at a height of
roughly $10\Rg$  \citep{2009ApJ...697.1164B}.

This height is suspiciously similar to the location of the stagnation
surface: the location at which the electromagnetically driven outflow
is arrested by the gravity of the black hole.  At present, the
stagnation surface is not consistently treated in simulations, which
often assume a sufficient number of charged particles to maintain the
electromagnetic currents required by the electrodynamics.  However,
the evacuation of this region (above by outflow, below by accretion)
should result in the creation of large electric fields, potentially
capable of both loading the jet with leptons and accelerating these
leptons to high energies. 

Acceleration of particles at gaps in black hole magnetosphere, in
analogy with particle acceleration in pulsar magnetosphere gaps, have
been discussed by a number of authors 
\citep[see, e.g.][]{1992SvA....36..642B,1998ApJ...497..563H,2011ApJ...730..123L},
including due to field-spin misalignment \citep{2007ApJ...671...85N}.
However, such studies are often either focused upon regions unassociated with
the stagnation surface, require energetic backgrounds, or ignore the
post-gap evolution of the resulting energetic particle population \citep{2011ApJ...730..123L},
which we find critical to reproducing the observed particle densities
in M87.  Existing efforts to numerically model the post-gap process
have not yet self-consistently modeled the gap itself
\citep{2010MNRAS.409.1183V}.  Thus, these are not immediately
applicable beyond the immediate scope of their derivation, and none
are directly relevant for the relativistic lepton loading of M87. 

Here we show that for M87 this is indeed expected to be the case.
Unscreened electric fields are capable of initiating an inverse-Compton
pair-creating cascade.  Within this process the number density of
nonthermal particles is a function primarily of the soft-photon
background produced by the surrounding accretion flow and within the
jet itself.  Most importantly, we show that for M87 the inferred
nonthermal lepton number densities match those implied by recent
mm-VLBI observations that resolve the jet launching region.
In \S\ref{sec:pcass} we discuss physical processes responsible for
particle acceleration and cooling in the gap near the
stagnation surface.  In \S\ref{sec:part-creat-accel}, we discuss the
properties of post-gap cascade and describe our numerical method.  In
\S\ref{M87SEC} and \S\ref{sec:application-sgr-a} we discuss our
results for M87 and Sgr~A*, respectively. In
\S\ref{sec:conclusions} we conclude.

\section{Particle Creation and Acceleration in the Stagnation Surface} \label{sec:pcass}

The stagnation surface is continually being evacuated as material
either accelerates outward in the jet or inward into the black hole.
In the absence of some mechanism to continually replenish the density
of charges the result is a charge-starved region, incapable of
screening the large electric fields that will be present.  The typical
electric field strength within the stagnation surface is set by the
angular velocity of the magnetic field line footprints, $\Omega_F$,
dragged about by the spinning black hole or the surrounding accretion
flow, 
\begin{equation}
E\approx\Omega_F R B/c,
\label{eq:ecorotation}
\end{equation}
where $R$ is the cylindrical radius of
the stagnation surface.  For black hole driven jets $\Omega_F$ is set,
in turn, by the angular velocity of the black hole horizon, 
\begin{equation}
\Omega_{\rm H}= ac/2r_{\rm H},\label{eq:omegaf}
\end{equation}
where 
$r_{\rm H} \equiv \Rg (1+\sqrt{1-a^2})$ is the size of the horizon in
Boyer-Lindquist coordinates.  Typically, analytical studies and 
GRMGHD simulations find
$\Omega_F\approx0.25\Omega_{\rm H}$--$0.5\Omega_{\rm H}$ \citep{1977MNRAS.179..433B,2004MNRAS.350.1431K,2010ApJ...711...50T,2012MNRAS.423.3083M,2013MNRAS.436.3741P,2014PhRvD..89b4041L}.

This unscreened electric field will necessarily accelerate electrons
across the charge-starved gap.  That is, given a gap thickness of
$\Delta$, in principle, the maximum electron Lorentz factor is 
\begin{equation}
\gamma_{\rm max} \approx e E \Delta.
\label{eq:gammamax}
\end{equation}
For gap sizes of order the
gravitational radius, in the case of M87 this
produces Lorentz factors of $10^{13}$, more than sufficient to produce
an inverse-Compton cascade.  In practice, $\gamma_{\rm max}$ is
limited by radiative losses and the size of the gap is set by the
cascade.  Here we estimate the gap thickness and corresponding pair
densities injected into the jet by the gap.

\begin{figure}
\begin{center}
\includegraphics[width=\columnwidth]{fig1.eps}
\end{center}
\caption{Schematic of a small portion of the stagnation surface and
  the associated gap region.  Within the gap particles are rapidly
  accelerated to asymptotic Lorentz factors, limited by inverse
  Compton cooling on the ambient soft-photon background (red
  photons), inducing a net charge separation.  The resulting
  up-scattered gamma rays (blue photons) pair produce on the
  soft-photon background, resulting in a pair catastrophe.  When the
  density in Equation (\ref{eq:ngap}) is reached, particle acceleration ceases,
  and the remaining post-gap cascade produces an asymptotic lepton
  density at fixed charge density.  Particles propagating downward are
  accreted by the black hole, while those propagating upward populate
  the black hole jet.  Due to the high Lorentz factors typically
  involved, the entire process is essentially one-dimensional; the
  oblique propagation angles of the resulting gamma rays are
  exaggerated for clarity.  Relevant locations and characteristic
  length scales are labeled (see the main text for details).}\label{fig:gap}
\end{figure}

\subsection{Radiative Losses}

\subsubsection{Inverse Compton}
The source of seed photons for inverse-Compton scattering is not
completely clear.  Nevertheless, we might assume that these are
related to the observed emission from the source.  Thus, if the
observed luminosity is synonymous with the seed photon luminosity we
will have 
\begin{equation}
L_s \approx \Omega r_s^2 u_s c/3 = \Omega D^2 F_{s}\,,
\label{eq:Ls}
\end{equation}
where $r_s$ is the size of the emission region, $\Omega$ is
the solid angle of the emission region (i.e., beaming factor), $u_s$
is the local energy density of seed photons, $D$ is the source
distance, and $F_{s}$ is the observed seed-photon flux.  Thus, in
terms of the observed flux,
\begin{equation}
u_s = 3 \frac{D^2 F_s}{r_s^2 c} = \frac{3 L_s}{4\pi r_s^2 c}\,,
\label{eq:us}
\end{equation}
where $L_s$ is the seed photon luminosity.
The cooling length to inverse-Compton scattering is related to
$u_s$ and $\gamma_{\rm max}$ via 
\begin{equation}
\ell_{\rm IC} = \frac{3 m_e c^2}{4\sigma_T u_s \gamma_{\rm max}}, 
\label{eq:ellIC}
\end{equation}
where
$\sigma_T$ is the Thomson cross section\footnote{This assumes that
$2\gamma_{\rm max}^2 \epsilon_s \lesssim \gamma_{\rm max} m_e c^2$,
which we will verify is approximately true for M87 in what follows.
If this is not true, $\ell_{\rm IC}$ is larger, producing a weaker
limit upon $\gamma_{\rm max}$.}.
This defines the typical distance over which particles can accelerate
within the gap, thus 
\begin{equation}
\gamma_{\rm max,IC} m_e c^2 \approx e E \ell_{\rm IC}\,,
\label{eq:gammamaxiccond}
\end{equation}
which implies a
limiting Lorentz factor of
\begin{equation}
\gamma_{\rm max,IC} = \left(\frac{3 e R\Omega_F B}{4\sigma_T u_s c}\right)^{1/2}
=
4.2\times10^8 R_{15}^{1/2} \Omega_{F,-4}^{1/2} B_2^{1/2} u_{s,0}^{-1/2}\,,
\label{eq:gammamaxIC}
\end{equation}
after inserting the expression for $E$ in which we have adopted the
standard notation, $R\equiv R_{15} 10^{15}~\cm$, 
$\Omega_F\equiv\Omega_{F,-4} 10^{-4}~\s^{-1}$, $B\equiv B_2 10^2~\G$,
and $u_s\equiv u_{s,0}~\erg~\cm^{-3}$.  The associated cooling length
is 
$\ell_{\rm IC}=2.2\times10^9 R_{15}^{-1/2} \Omega_{F,-4}^{-1/2}B_2^{-1/2} u_{s,0}^{-1/2}~\cm$.

\subsubsection{Curvature}
Curvature radiation losses provide a second, seed-photon independent,
limit upon $\gamma_{\rm max}$.  The power emitted depends upon the
curvature radius of the magnetic field, $r_c\approx R$, and is given
by  \citep{2011ApJ...730..123L}
\begin{equation}
P_{\rm C} = \frac{2 e^2 c \gamma_{\rm max}^4}{3R^2}
\label{eq:pcurv}
\end{equation}
The
corresponding cooling length is 
\begin{equation}
\ell_{\rm C} = \frac{3 R^2 m_e c^2}{2e^2\gamma_{\rm max}^3},
\label{eq:ellc}
\end{equation}
implying a
limit of
\begin{equation}
\gamma_{\rm max,C} = \left(\frac{3 R^3 \Omega_F B}{2 e c}\right)^{1/4}
=
3.2\times10^{10} R_{15}^{3/4} \Omega_{F,-4}^{1/4} B_2^{1/4}\,.
\label{eq:gammamaxc}
\end{equation}
In practice, the maximum particle Lorentz factor is set by the
minimum radiative loss limit.  As we will see in Section \ref{M87SEC} the
limit arising due to inverse-Compton losses is typically more severe.
This is true if
\begin{multline}
u_s > \left( \frac{3 e^3 \Omega_F B}{8 \sigma_T^2 R c}\right)^{1/2}
=
1.8\times10^{-4} \Omega_{F,-4}^{1/2} B_2^{1/2} R_{15}^{-1/2}~\erg~\cm^{-3}\\
\quad\Rightarrow\quad
F_s \gtrsim 
1.8\, r_{s,16}^2 D_{26}^{-2} \Omega_{F,-4}^{1/2} B_2^{1/2} R_{15}^{-1/2}~\mJy~\THz
\,,
\end{multline}
where $r_s\equiv r_{s,16}10^{16}~\cm$ and 
$D\equiv D_{26}10^{26}~\cm$.  That is, as long as the seed photon
density is sufficiently high, the gap dynamics is set by
inverse-Compton losses.  Even for the highly sub-Eddington sources
(e.g., M87 and Sgr~A*) this is often satisfied.

\subsubsection{Synchrotron}
Generally the electric field in the gap will {\em not} be aligned with
the magnetic field.  As a consequence, accelerated particles will
gyrate, and thus emit synchrotron radiation.  This can provide an
alternative source of hard photons oriented at substantial angles to
the magnetic field.  The cooling length for synchrotron is 
$\ell_{\rm S} = \ell_{\rm IC} u_s/u_B$, where $u_B=B^2/8\pi$, and thus
we generally have $\ell_{\rm S}\ll\ell_{\rm IC}$, since the seed
photon energy density is typically much smaller than the magnetic
energy density in the sources of interest.  However, note that
synchrotron is only capable of cooling the {\em transverse} motion of
the particles.  While this serves to ensure in extremely short time 
that the leptons are traveling {\em along} magnetic field
lines (making the particle acceleration essentially a one-dimensional
process), it does not otherwise limit the particle Lorentz factors.

\subsubsection{Collective Processes}
Finally, we consider the importance of collective effects due to the
counter-streaming electron-positron plasmas.  Due to their opposite
charges, electrons and positrons will be accelerated in opposite
directions, producing a pair of cold, counter-propagating lepton
beams.  This is unstable to the well-known two-stream instability,
modified slightly due to the opposite charges in the two beams (which
we refer to as the ``counter-streaming instability'' and presented
briefly in Appendix \ref{app:csts}).  The instability growth rate 
is generally dependent upon the stream densities, corresponding to a
cooling length of
\begin{equation}
\ell_{\rm CSI} 
\approx 
\left(\frac{2 m_e \gamma_{\rm max}^3}{\pi e^2 n_{\rm g}}\right)^{1/2} c\,,
\end{equation}
and thus we cannot fully evaluate the importance of plasma
instabilities without an estimate of the pair densities.  However, for
the estimate presented below (Equation \ref{eq:ngap}),
$\ell_{\rm CSI} \approx 7.1\times10^{18} u_{s,0}^{-1} \eta_{\rm th}^{1/4}~\cm$,
and therefore this is never relevant.

\subsection{Pair Density within the Gap} \label{sec:gapdensity}
As the limiting radiative process, inverse-Compton cooling also
provides a limit upon the gap thickness.  The up-scattered photons have
energies of roughly 
$\min(2\gamma_{\rm max}^2 \epsilon_s, \gamma_{\rm max} m_e c^2)$.
Thus, at meV energies, comparable to the photon energy near the peak of
the spectral energy distribution (SED) of the sources of primary
interest here (see Sections~\ref{sec:seed-photon-density-m87} and
\ref{sec:seed-photon-density-sgr}), the up-scattered photon energy is similar to that of
the incident electron.  Thus, the up-scattered photons will pair
produce off of photons with energies above
\begin{equation}
\epsilon_{\rm th}
\approx
\frac{4 m_e c^2}{\gamma_{\rm max}}
=
4.8\, R_{15}^{-1/2} \Omega_{F,-4}^{-1/2} B_2^{-1/2} u_{s,0}^{1/2}~\meV\,,
\end{equation}
and thus just above the SED peak for our fiducial parameters.
Hence, the inverse-Compton cooling of the gap-accelerated pairs
initiates a pair-production catastrophe.

The pair-production cross section peaks near $2 \epsilon_{\rm th}$ at
$\sigma_{\gamma\gamma}\approx 0.35\sigma_T$ \citep{Goul-Schr:67}.
Thus, the mean free path of the up-scattered seed photons is
approximately
\begin{equation}
\ell_{\gamma\gamma} \approx \frac{2 \epsilon_{\rm th}}{u_{s,\rm th} \sigma_{\gamma\gamma}}
\approx 6.6\times10^{10} R_{15}^{-1/2} \Omega_{F,-4}^{-1/2} B_2^{-1/2}
u_{s,0}^{-1/2} \eta_{\rm th}
~\cm\,,
\label{eq:ellgg}
\end{equation}
where $u_{s,\rm th}\equiv u_{s,0}/\eta_{\rm th}$ is the energy density
of seed photons with energies above $\epsilon_{\rm th}$.  For a source
with infrared spectral index $\alpha$ (i.e.,
$F_\nu\propto\nu^{-\alpha}$, in M87 $\alpha\approx1.2$), the number of
seed photons above an energy $\epsilon_s$ is 
$\propto \epsilon_s^{-\alpha}$.  Thus, if there is a spectral break at
$\epsilon_b$, below which the seed photon energy density may be
neglected,
\begin{equation}
\eta_{\rm th} 
=
\begin{cases}
1 & \epsilon_{\rm th}<\epsilon_b\\
(\epsilon_{\rm th}/\epsilon_b)^{\alpha} & \epsilon_{\rm th}\ge \epsilon_b\,.
\end{cases}
\label{eq:etath}
\end{equation}
For sufficiently high $\epsilon_{\rm th}$, it is possible for
$\eta_{\rm th}\gg1$, implying that 
$\ell_{\gamma\gamma}\gg\ell_{\rm IC}$.  When $\eta_{\rm th}=1$,
this is larger than $\ell_{\rm IC}$ by roughly a factor of three, 
implying in this case that the gap scale height comparable to either.
A more quantitative model of the gap lepton population may be obtained
in one-dimension by assuming that the pairs are accelerated
instantaneously to their asymptotic velocities. As shown in Appendix
\ref{app:1dgap}, this implies a gap thickness of
\begin{equation}
\Delta  \approx \sqrt{2\ell_{\rm IC}\ell_{\gamma\gamma}}
\approx
1.7\times10^{10}  R_{15}^{-1/2} \Omega_{F,-4}^{-1/2} B_2^{-1/2}
u_{s,0}^{-1/2} \eta_{\rm th}^{1/2}
~\cm\,,
\label{eq:gapthickness}
\end{equation}
consistent with this.  Generally, $\ell_{\gamma\gamma}$ is larger than
$\ell_{\rm IC}$ and therefore the asymptotic Lorentz factor is gained
on a length scale $\ell_{\rm IC}\ll \Delta$ thereby justifying our
assumption that the pairs are instantaneously accelerated.

Counter-propagating positrons or electrons constantly reinitiate the
cascade, which continues until sufficient charges are produced to
screen the electric fields.  Thus we would anticipate that the gap can
be evacuated only over a single scale height.  The top of the gap is
then roughly defined when the charge density is sufficient to
generate the gap electric field gradients, and thus the lepton density
in the lab frame can be expressed as (see Appendix~\ref{app:1dgap}):
\begin{equation}
n_{\rm g} = \frac{\bmath{\nabla}\cdot\bmath{E}}{4\pi e} \approx \frac{E}{4\pi e \Delta}
=
3.3\, R_{15}^{3/2} \Omega_{F,-4}^{3/2} B_{2}^{3/2}
u_{s,0}^{1/2}
\eta_{\rm th}^{-1/2}
~\cm^{-3}\,.
\label{eq:ngap}
\end{equation}
Note that this exceeds the Goldreich-Julian density by a
\emph{multiplicity} factor given by
\begin{equation}
  \label{eq:multiplicity}
  \eta = \frac{R}{\Delta},
\end{equation}
resulting from the smaller typical scale set by the gap
thickness.\footnote{This factor was not included in previous
works (e.g., \citealt{2011ApJ...730..123L,2011ApJ...735....9M}).} Here
we have assumed that only a single species is present at the top of
the gap, an assumption that is justified by the rapid acceleration of
electrons and positrons in opposite directions. However, the nonlinear
response of the magnetosphere will supply the charges of the opposite
sign to screen the gap electric field and ensure that the
microphysical current matches the global magnetospheric current (which
is generically of order of Goldreich-Julian current, $j_{\rm GJ} = e n_{\rm GJ}c$).

We note that first-principles particle-in-cell (PIC) simulations are
needed in order to simulate the cascade in detail and properly compute
the flow of charges and connection between the pair cascade and global
MHD currents. Fortunately, our simplified, steady-state consideration
of the gap structure appears to capture the most crucial aspects of
the gap structure to within an order of magnitude. For instance, in
PIC simulations of polar cascades in pulsar magnetosphere 
\citep[hereafter,
\citetalias{2013MNRAS.429...20T}]{2013MNRAS.429...20T}, the simulated
pair multiplicity $\eta\sim100$ (for their polar gap thickness of
$\Delta\sim1.7\times10^4$~cm and a characteristic length scale
$R=10^6$~cm; see, e.g., Figure~22 in \citetalias{2013MNRAS.429...20T})
agrees with our estimate \eqref{eq:multiplicity}, which gives
$\eta = R/\Delta\approx60$. Furthermore, in agreement with the findings of
\citetalias{2013MNRAS.429...20T}, our gap thickness is much larger
than the skin depth,
\begin{equation*}
\frac{\Delta}{d_e}
=
5.6\times10^{5} \eta_{\rm th}^{1/4} R_{15}^{1/4} \Omega_{F,-4}^{1/4}
B_2^{1/4} u_{s,0}^{-1/4}\,,
\end{equation*}
with the skin depth given by
\begin{equation*}
d_e = \sqrt{ \frac{m c^2}{4\pi e^2 n_g} }\,.
\end{equation*}
Encouraged by this agreement, we will adopt the results of our steady
state gap model and consider astrophysical consequences.

The relativistic outflow of particles at the top of the gap can carry
a substantial total kinetic luminosity.  Given the above estimates
for the number density and typical Lorentz factor, the kinetic flux in
the outflowing leptons is
\begin{equation}
\begin{aligned}
F_{{\rm lep}} &= \gamma_{\rm max,IC} m_e c^3 n_{\rm g}\\
&=
3.4\times10^{13}
R_{15}^2 \Omega_{F,-4}^2 B_2^2
\eta_{\rm th}^{-1/2}
~\erg~\cm^{-2}~\s^{-1}\,.
\end{aligned}
\label{eq:felep}
\end{equation}
Integrating this across the entire stagnation surface gives a total
kinetic luminosity of
\begin{equation}
\begin{aligned}
L_{{\rm cascade}}
&\approx
\int_0^R 4\pi R' dR' F_{{\rm lep}}\\
&=
1.1\times10^{44}
R_{15}^4 \Omega_{F,-4}^2 B_2^2
\eta_{\rm th}^{-1/2}
~\erg~\s^{-1}\,.
\end{aligned}
\label{eq:Llep}
\end{equation}
Note that typically $\Omega_F\propto M^{-1}$ and $R\propto M$, where
$M$ is the mass of the black hole.  Hence, at fixed magnetic field
strength the pair luminosity scales as $M^2$, implying that
$L_{e^{\rm lep}}/L_{\rm Edd}\propto M$.  Our fiducial numbers correspond
roughly to M87, and thus a black hole  mass of order
$10^{10}~M_\odot$, for which the above luminosity is about
$3\times10^{-4}L_{\rm Edd}$. This suppresses the cascade luminosity for
less massive black holes, as we will see on an example of Sgr~A* in 
\S\ref{sec:application-sgr-a}.

\subsection{Gap Stability}
\begin{figure}
\begin{center}
\includegraphics[width=\columnwidth]{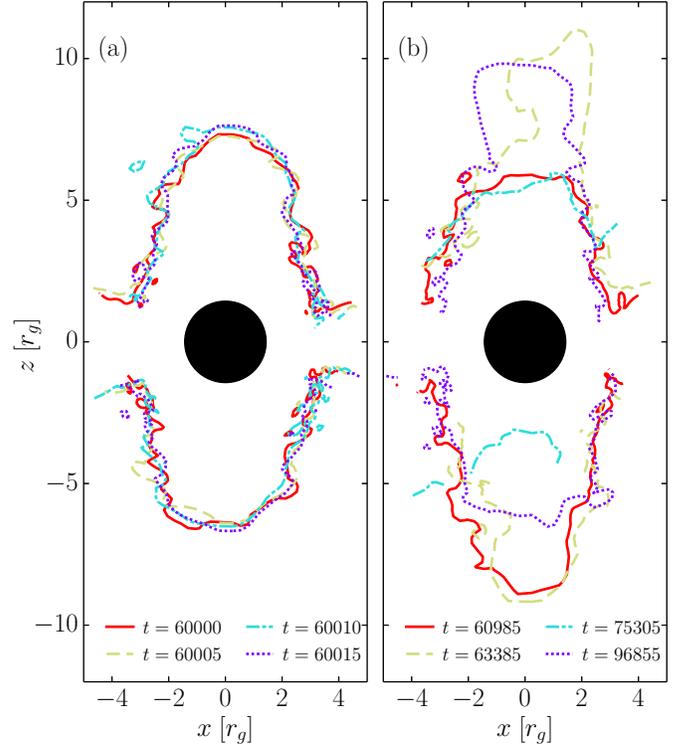}
\end{center}
\caption{Meridional ($x{-}z$) snapshots of the stagnation surface, at
  which the radial velocity vanishes ($u^r = 0$), shown at
  different times with different colors (see legend) for a simulation
  of magnetically-arrested disk around a spinning black hole with
  dimensionless spin $a = 0.9$ (Tchekhovskoy et al., in
  preparation).  For clarity, we only show the highly magnetized
  region, $p_{\rm mag}/\rho c^2 > 10$. The black
  hole event horizon is shown as the black filled circle. Panel (a)
  shows a series of snapshots of the stagnation surface taken at
  intervals comparable to the light crossing time, $\Delta t\approx 5\Rg/c$. The
  stagnation surface is clearly variable on such a short timescale,
  with its radial position changing by $5{-}10$\%. On average, the
  stagnation surface is at a distance $r \approx 5{-}10\Rg$. Panel (b) shows
  how the shape of the stagnation surface evolves on longer time
  scales, $\Delta t\approx10^4 \Rg/c$. On such longer time scales the 
  amplitude of variability can be as high as $50$\%.  For example,
  the shape of the surface can be substantially
  distorted by infalling gas that pushes the jet (e.g., see cyan
  dash-dotted line, which corresponds to $t = 75305\Rg/c$).
}\label{fig:stagsurface}
\end{figure}

The lepton structure in the gap is generally quite unstable, both due
to the large-scale GRMHD processes that lead to its formation and the
particle acceleration that fill it.  In the case of its generic
structure, the typical variability timescale is of order the light
crossing time of the radial position of the gap, typically
$\approx10\Rg$. The radial position of the gap varies by about $\lesssim10$\% on
this time scale, as can be seen in Figure~\ref{fig:stagsurface}.

The particle acceleration processes will induce local variability on
much shorter timescales, comparable to the light crossing time across
the gap (see Appendix \ref{app:1dgapdyn}).  In our fiducial model,
this is roughly 
$0.7 R_{15}^{-1/2} \Omega_{F,-4}^{-1/2} B_2^{-1/2} u_{s,0}^{-1/2}~\s$.
For M87, this is roughly six orders of magnitude smaller than
$GM/c^3$!  However, in practice, the gap has a transverse extent that
greatly exceeds $\Delta$, and therefore usually many independently
fluctuating regions are visible at once.  Assuming that each
independent region has a local radius of $\Delta$, if the visible 
extent of the stagnation surface is $R_V$, the net degree of
variability on the particle variation timescale is roughly 
\begin{equation}
\frac{\delta n_{\rm g}}{\langle n_{\rm g}\rangle} \approx \frac{\Delta}{R_V}\,.
\end{equation}

For low frequencies, $R_V\approx R$, and this reduction is enormous.
However, for the very-high energy gamma rays produced directly by
inverse Compton scattering, the intrinsic relativistic beaming will
substantially reduce $R_V$.  For emission associated with electrons
with Lorentz factor $\gamma$, a patch of the surface that will be
visible has a typical size of $R/\gamma$, implying that for
$\gamma\gtrsim 10^6$ the observed particle density variations, and
therefore the flux variations, will be of order unity on the particle
acceleration timescale.  In our fiducial model, where the soft seed
photon energy density peaks around 1~meV, this corresponds to a
gamma-ray energy above a few GeV.  

As the energy $\epsilon_\gamma$ of a gamma-ray increases, the number
density of soft photons that lie above the pair creation threshold
increases, and so does the optical depth $\tau_{\gamma\gamma}$ to pair
production.  The existence of the pair catastrophe implies that at
sufficiently high $\epsilon_\gamma$ the optical depth to pair
production is high, i.e., $\tau_{\gamma\gamma}\gg1$.  Since
the energy threshold for annihilation of a gamma ray is
$\propto\epsilon_\gamma^{-1}$, the number density of soft photons
above the pair creation threshold (and therefore the optical depth to
annihilation) scales as $\propto\epsilon_\gamma^{\alpha}$.   The
normalization of the total gamma-ray optical depth can be obtained in
terms of $\ell_{\gamma\gamma}$, set by $\tau_{\gamma\gamma}=1$ for
gamma rays with energy $\gamma_{\rm max,IC} m_e c^2$:
\begin{equation}
\tau_{\gamma\gamma}(\epsilon_\gamma) \approx \frac{r}{\ell_{\gamma\gamma}}
\frac{n_s(\epsilon_s>4 m_e^2 c^4/\epsilon_\gamma)}
     {n_s(\epsilon_s>4 m_e c^2/\gamma_{\rm max,IC})}
\approx
\frac{r}{\ell_{\gamma\gamma}} \left(\frac{\epsilon/m_e c^2}{\gamma_{\rm max,IC} }\right)^\alpha\,.
\end{equation}
Hence, for gamma-rays with energies
\begin{equation}
\epsilon_\gamma
\lesssim\epsilon_{\gamma,\rm min} \equiv 
\gamma_{\rm max, IC} m_e c^2 \left(\frac{\ell_{\gamma\gamma}}{r}\right)^{1/\alpha}\,,
\label{eq:glim}
\end{equation}
the seed photon distribution will be optically thin to the resulting inverse
Compton gamma rays, allowing the latter to escape.  For values typical of M87, this corresponds to
energies of order $\epsilon_\gamma\approx630$~GeV, above which significant
attenuation can be expected.

\section{Particle Creation and Acceleration Beyond the Stagnation Surface}\label{sec:part-creat-accel}
The leptons streaming out of the gap opened near the stagnation
surface will still have near-maximal Lorentz factors, 
$\gamma_{\rm max,IC}$.  As a result, they will continue to pair
produce beyond the gap each time they traverse a distance of roughly $\ell_{\rm IC}$.
However, unlike the results of the previous section, since the leptons are
outside of the gap, they are no longer rapidly re-accelerated, and thus
the pair catastrophe is no longer being driven.  The result is a
redistribution of the energy of gap-injected leptons among a set of
daughter leptons produced via the resulting final cascade.

\subsection{Qualitative Asymptotic Estimates}

We have already noted that for gamma rays with energies below 
a lower limit determined by the seed photon SED the ambient soft
photon population will be transparent, halting the pair production
cascade.  
Explicitly, for a power law seed photon distribution above a
low-energy cutoff $\epsilon_b$ (as described above Equation
\ref{eq:etath}) the fraction of gamma rays with energies above
$\epsilon_{\gamma,\rm min}$ following a single scattering is
\begin{equation}
f_{>\epsilon_{\gamma,\rm min}}(\gamma)\approx
\begin{cases}
\left(\epsilon_{\gamma,\rm min}/2\gamma^2\epsilon_b\right)^{-\alpha}
& \gamma\le \sqrt{\epsilon_{\gamma,\rm min}/2\epsilon_b}\\
1 & \text{otherwise,}
\end{cases}
\end{equation}
where we have assumed that $\epsilon_b$ is sufficiently small.  This is
necessarily a strong function of the initial lepton Lorentz factor,
dropping rapidly for Lorentz factors below 
$\sqrt{\epsilon_{\gamma,\rm min}/2\epsilon_b}$, and thus
$\epsilon_{\gamma,\rm min}$ implies a limit upon $\gamma$.

However, since the energies of the upscattered gamma rays are
typically small in comparison to the lepton energy near this limit,
i.e., $2\gamma^2\epsilon_b \ll \gamma m c^2$, in practice the lepton
will undergo many scatterings before cooling appreciably.  That is,
there will be roughly 
\begin{equation}
N_{\rm scat} \approx \frac{\gamma m c^2}{2\gamma^2 \epsilon_b} = \frac{m c^2}{2\gamma \epsilon_b}\,,
\end{equation}
opportunities to produce a gamma ray with energy above
$\epsilon_{\gamma,\rm min}$.  This increases with decreasing $\gamma$
and $\epsilon_b$, ameliorating the decrease in
$f_{>\epsilon_{\gamma,\rm min}}$ due to the latter. Thus the number of
pair producing gamma rays is expected to be approximately 
$N_{\rm scat} f_{>\epsilon_{\gamma,\rm min}}$.  Setting this number to unity
provides the desired limit upon the lepton Lorentz factor, below which
electrons cease to typically have sufficient energy to continue the
pair cascade:
\begin{equation}
\gamma_{\gamma\gamma} \approx \left( 
 \frac{2^{1-\alpha}\epsilon_b^{1-\alpha} \epsilon_{\gamma,\rm min}^{\alpha}}{m c^2} 
\right)^{1/(2\alpha-1)}\,.
\label{eq:ggg}
\end{equation}
For our fiducial model, 
$\gamma_{\gamma\gamma}\approx7.7\times10^4$.
This limit on the
redistribution of energy implies an associated limit on increase in
the particle density.  An even redistribution, maximizing the number
of leptons produced, results in
\begin{equation}
\frac{n_\infty}{n_{\rm g}} \approx \frac{\gamma_{\rm max,IC}}{\gamma_{\gamma\gamma}}\,,
\label{eq:enhancement}
\end{equation}
and is roughly $6000$ daughter leptons for each lepton produced by the
gap for our fiducial model.

The absence of the generation of energetic gamma rays does not imply
the absence of additional inverse-Compton cooling.  Hence, a second,
typically more stringent limit arises from inverse-Compton cooling
directly.  For the cooling length to become comparable to the scale
height of the seed photon distribution requires
\begin{equation}
\frac{3m_e c^2}{4\sigma_T u_s \gamma_\infty}\gtrsim r
\quad\Rightarrow\quad
\gamma_\infty \lesssim \gamma_{\rm max,IC} \frac{\ell_{\rm IC}}{r}
\approx 10^2 u_{s,0}^{-1} r_{16}^{-1}\,.
\label{eq:gammainfty}
\end{equation}
We note that in our fiducial model, the cooling of the leptons from
$\gamma_{\gamma\gamma}$ to $\gamma_\infty$ is unable to produce
gamma rays energetic enough to pair produce. Therefore, the resulting
radiation can escape and can be a substantial
source of emission.

Since $\gamma_\infty n_\infty \ll\gamma_{\rm max,IC} n_{\rm g}$ most of the
energy imparted to the leptons by the accelerating electric field across the gap has been
radiated via the Comptonization of the soft seed photons.  Typical
energies range from the optical to the GeV, with an SED that depends
upon that of the underlying soft photon population.  However, this
emission will be highly beamed along the magnetic field lines that
govern the particle motions.

\subsection{Numerical Asymptotic Estimates} \label{sec:NAE}

In practice the estimates obtained in the previous section depend
critically upon assumptions regarding the final electron spectrum,
that is itself dependent upon the soft photon SED.  For this reason we
also computed the asymptotic number and energy distribution of
electrons produced due to inverse-Compton scattering off of a given
seed photon SED from a nearly mono-energetic injection at high Lorentz
factor.  Throughout this section, for compactness we will measure
energies in units of the electron rest mass energy.

We make two simplifying assumptions, both of which are
almost certainly well justified.  The first is that we may ignore any
subsequent pair annihilation due the comparatively high lepton
energies and low lepton densities.  The second is that the number
density of soft seed photons vastly exceeds the number densities of
up-scattered gamma rays, a consequence of the short mean free path to
pair production and the low lepton densities.  These remove the
non-linear terms in the coupled Boltzmann equations for the leptons
and the gamma-rays.  Thus, it is sufficient to consider the evolution
of a single lepton through multiple Compton generations.  That is, we
compute the energy probability distribution of a single lepton as a
function of the number of scatterings instead of time.  The asymptotic
form of this probability distribution then describes the asymptotic
spectrum of a full population of leptons.

Explicit inputs are the seed photon spectrum and electron injection
energy, which we leave as a parameter, $\epsilon_0$.
For the former we assume a power law model between some minimum
($\epsilon_m$) and maximum ($\epsilon_M$) energies, consistent with
the observed radio and infrared emission from the sources of interest,
i.e.,
\begin{equation}
f_s(\epsilon)\equiv \frac{{\rm d}N_s}{{\rm d}\epsilon}\propto
\begin{cases}
\epsilon^{-\alpha-1} & \epsilon_m < \epsilon < \epsilon_M\\
0 & \text{otherwise.}
\end{cases}
\label{eq:spd}
\end{equation}
The values of $\epsilon_m$ and $\epsilon_M$ are determined
by the particular source under consideration, though typically when a
large enough dynamic range is chosen the resulting evolution becomes
insensitive to their specific choice.\footnote{The reason is that
when $\epsilon_m\ll4m_e^2c^4/\epsilon_0$ the up-scattered photons will
be unable to pair produce (and the energy loss becomes comparatively
insignificant) and when 
$\epsilon_M\gg4m_e^2c^4/\epsilon_0$ the number density of photons with
energies near $\epsilon_M$ will be small in comparison to those
responsible for the bulk of the particle creation (and thus may be
neglected).}  The linearity in the gamma-ray distribution is synonymous
with assuming that $f_s$ remains fixed throughout the post-gap pair
cascade.

With the above we may now compute the update in the lepton energy
distribution with each scattering.  We do this in two steps, first
computing the implied up-scattered gamma-ray distribution associated
with a mono-energetic electron distribution, and then from that
inferring the changes to the electron distribution arising from
inverse-Compton losses and pair production.  To compute the former, we
begin with the relationship between the original seed photon energy
$\epsilon_s$, up-scattered gamma-ray energy $\epsilon_\gamma$, and
initial electron energy $\epsilon_{e,i}$:
\begin{equation}
\epsilon_s(\epsilon_\gamma,\epsilon_{e,i})
=
\begin{cases}
\displaystyle
\frac{\epsilon_\gamma}{2\epsilon_{e,i}\left(\epsilon_{e,i}-\epsilon_\gamma\right)}
& \epsilon_\gamma<\epsilon_{e,i}\\
0 & \text{otherwise,}
\end{cases}
\label{eq:es}
\end{equation}
where we assumed $\epsilon_s\ll \epsilon_{e,i}$ and approximated energy of the electron
post-collision as $\epsilon_{e,f} = \epsilon_{e,i}-\epsilon_\gamma$.  Equation
\eqref{eq:es} may be obtained from the standard Compton scattering formula in
the high-energy electron and oblique seed photon limits.  As a result,
for a monoenergetic electron distribution, the up-scattered gamma-ray
distribution is
\begin{equation}
g(\epsilon_\gamma,\epsilon_{e,i})
=
\frac{d\epsilon_s}{d\epsilon_\gamma}
f_s[\epsilon_s(\epsilon_\gamma,\epsilon_{e,i})]
=
\frac{f_s[\epsilon_s(\epsilon_\gamma,\epsilon_{e,i})]}
{2(\epsilon_{e,i}-\epsilon_\gamma)^2}\,.
\end{equation}

As mentioned above, since $\epsilon_s\ll\epsilon_{e,i},\epsilon_\gamma$, we approximate
the electron energy following scattering by
$\epsilon_{e,f}\approx\epsilon_{e,i}-\epsilon_\gamma$. Therefore, if
$f_j(\epsilon_e)$ is the electron distribution after the $j$th
scatter, the electron distribution is rearranged by inverse-Compton
losses to
\begin{equation}
f^{\rm lep}_{{\rm IC},j+1}(\epsilon_e)
=
\int_0^\infty d\epsilon_{e,i} g(\epsilon_{e,i}-\epsilon_e,\epsilon_{e,i}) 
f^{\rm lep}_j(\epsilon_{e,i})\,.
\label{eq:lepic}
\end{equation}
In practice the integral need only be computed up to the initial
maximum energy of the electron distribution.

Similarly, since $\epsilon_\gamma\gg2\gg\epsilon_s$, i.e., much greater
than the rest mass of the produced pair, which is itself much greater
than that of the seed photons, we may approximate the energy of each
lepton in
the resulting pair by $\epsilon_e=\epsilon_\gamma/2$.  Hence, the distribution of
new pairs is 
\begin{equation}
\label{eq:lepgg}
f^{\rm lep}_{\gamma\gamma,j+1}(\epsilon_e)
=
4\int_0^\infty d\epsilon_{e,i}
g(2 \epsilon_e,\epsilon_{e,i}) f_j^{\rm lep}(\epsilon_{e,i})
\ThetaFunc(2\epsilon_e-\epsilon_{\gamma,\rm min})
\,,
\end{equation}
where $\ThetaFunc(x)$ is the Heaviside function and ensures that the
integral is only over regions in which the gamma-ray can pair produce
on the soft background.  Note that we assume that all gamma-rays will
pair produce as long as their energy exceeds the threshold
$\epsilon_{\gamma,\rm min}$, above which the seed photon distribution
is optically-thin to pair production.

In principle, we should then set $\epsilon_M$ by the
pair production optical depth; from Equation (\ref{eq:glim}) this
gives a typical 
\begin{equation}
\epsilon_M\approx 4m_e^2 c^4/(630~\GeV)
\label{eq:epsilonM}
\end{equation}
which for our fiducial parameters gives $\epsilon_M\approx1.6~\eV$. In
practice, we find the simulation results are insensitive the
particular value we choose for $\epsilon_{\rm M}$.  In all of our
simulations, unless specified otherwise, we choose $\epsilon_{M} =
0.8$~eV (as we discuss below, results for $\epsilon_{M} =
1.6$~eV are essentially identical; see Figure~\ref{fig:nvse0}).
In a similar fashion to Equation \eqref{eq:lepgg}, we can write down the distribution
function of photons that are below the pair production threshold,
$\epsilon_\gamma < \epsilon_{\gamma,\rm min}$:
these photons escape the system without undergoing scattering and thus
form the high-energy emission spectrum,
\begin{equation}
\label{eq:phgg}
f^{\rm ph}_{\gamma\gamma,j+1}(\epsilon_\gamma)
=
4\int_0^\infty d\epsilon_{e,i}
g(\epsilon_\gamma,\epsilon_{e,i}) f_j^{\rm lep}(\epsilon_{e,i})
\ThetaFunc(\epsilon_{\gamma,\rm min}-\epsilon_\gamma)
\,.
\end{equation}

Combining the above, we obtain the energy distribution of the leptons
and photons
in the $(j+1)$th generation via
\begin{align}
f^{\rm lep}_{j+1}(\epsilon_e) 
&= 
f^{\rm lep}_{{\rm IC},j+1}(\epsilon_e) + f^{\rm lep}_{\gamma\gamma,j+1}(\epsilon_e)
\,,
\label{eq:lep}\\
f^{\rm ph}_{j+1}(\epsilon_\gamma)  &= f^{\rm ph}_{j}(\epsilon_\gamma)
+ f^{\rm ph}_{\gamma\gamma,j+1}(\epsilon_\gamma)\, . \label{eq:ph}
\end{align}
Beginning with some initial injection distribution, we can compute the
asymptotic distribution, in which generations correspond in a loose
sense to time or height from the injection point.

For this, we integrate Equations~\eqref{eq:lep}--\eqref{eq:ph}
numerically.  We start the integration at the top of the gap, and
inject a single electron at an energy 
\begin{equation}
E_0 = \gamma_{\rm max,IC}m_ec^2,  \label{eq:E0}
\end{equation}
which is the energy gained by an electron between its
consequent encounters of seed photons in the gap and is therefore the
characteristic energy with which the electrons emerge from the gap.  
We discretize Equations~\eqref{eq:lep}--\eqref{eq:ph} on a logarithmic grid in energy,
which extends from $E_{\rm min} = 10^{-6}m_e c^2$ to
$E_{\rm max} = 2E_0$. For numerical reasons, we smooth out the initial
energy distribution by $1.5\%$ so it is properly represented on our
numerical grid, by choosing the following initial injection
distribution,
\begin{equation}
f^{\rm lep}_0(\epsilon_e) \equiv \frac{{\rm d}N_e}{{\rm d}\epsilon_e}=
\frac{1}{\sqrt{2\pi\sigma_E^2}}\exp\left[-\frac{(\epsilon_e-E_0)^2}{2\sigma_E^2}\right], 
\label{eq:flepinitial}
\end{equation}
with $\sigma_E = 0.015E_0$.

In order to ensure robust evolution of pair cascade numerically, 
we found it to be important to ensure lepton number
conservation under Inverse Compton cooling (Equation~\eqref{eq:lepic}). To
do this, we compute the discretization of the integral in
Equation~\eqref{eq:lepic} on two energy grids that contain the same number
of grid cells but are shifted relative to each other by half a
cell. If the changes in the number of leptons due to these two
discretizations are different by more than $50$\% relative to each
other, then we use their weighted sum that preserves the total number
of leptons exactly. Otherwise, we use the discretization that results
in the smallest change in lepton number. We implemented the numerical
code in a Cython-based Python library. In order to speed up the
integrations, we parallelized the code using OpenMP. We describe the numerical
results of cascade evolution in Section~\ref{sec:post-gap-cascade}.

\section{Application to M87}\label{M87SEC}

Thus far we have remained agnostic regarding the particular
sources for which particle acceleration at the stagnation surface is
relevant.  Here we apply this model to the generation of the
nonthermal particles in the radio jet of M87.  We begin with a short
overview of the relevant source parameters and their corresponding gap
parameters.

\subsection{M87 Source Parameters}
Observations of stellar dynamics within M87's sphere of influence
produce an inferred mass of $6.6\times10^9\,\Ms$, assuming a distance
of $17.9\,\Mpc$ \citep{2011ApJ...729..119G}.\footnote{While quantitative
differences appear if we were to adopt the lower mass of
$3.5\times10^9\,\Ms$ implied by gas dynamical modeling
\citep[][]{2013ApJ...770...86W}, the qualitative consequences remain
unchanged.}  The energy output of M87 peaks at 1~mm (300~GHz or 1~meV),
at which point the flux at Earth is roughly $1\,\Jy$.  The typical
isotropic-equivalent total luminosity is $L \approx
10^{42}~\erg~\s^{-1} \approx 10^{-6} L_{\rm Edd}$, where $L_{\rm Edd}$
is the Eddington luminosity. This makes M87 quite
underluminous and dominated by emission at wavelengths about the
peak frequency, as we discuss below.  The typical scale near the black hole is the gravitational
radius of the black hole, $\Rg\simeq 10^{15}\,\cm$.  
Here we discuss the empirical implications of various additional
observations for the physical properties near the stagnation surface.

\subsubsection{Jet Velocity and Orientation}

Apparent motions near the black hole of roughly $2c$ have been
observed \citep[e.g.,][]{2008JPhCS.131a2053W}.  These imply a minimum
Lorentz factor of $\Gamma\approx2.3$ by distances of $0.3\,{\rm mas}$,
corresponding to linear distances of 
$8\times10^{16}\,\cm\approx 80\Rg$, and implying an inclination of
$25^\circ$.   On scales of $1''$, more than three orders of magnitude
farther out, apparent motions of $6c$ have been observed, implying
$\Gamma>6$ at an inclination of $\approx10^{\circ}$.  Here we will
employ the radio implied Lorentz factor and associated orientation,
$\Gamma=2.3$ and $\Theta=25^\circ$, where specific values are required.

\subsubsection{Magnetic Field Strength Estimates}
The observational implications for the magnetic field in M87 depend
upon the structure and dynamics of the material in the vicinity of the
black hole.  Here we present two estimates, assuming in the first that
the observed mm-wavelength emission is due to a radiatively
inefficient accretion flow (RIAF, \citealt{1994ApJ...428L..13N,1995ApJ...452..710N}), and in the second that the
jet is Poynting dominated and responsible for the power observed in
the radio lobes.  Both reach similar conclusions regarding the field
strength.

As suggested by their name, RIAFs are characterized by their
extraordinarily low radiative efficiency, i.e., $L = \eta_d \dot{M}c^2$
with $\eta_d\approx10^{-3}$--$10^{-4}$.  Since most of the gravitational
binding energy is released, and therefore most of the luminosity is
produced, near the black hole, this implies an accretion flow density of
\begin{equation}
\rho 
\approx \frac{\dot{M}}{4\pi\Rg^2 \beta c} 
= \frac{L}{4\pi \eta_d \Rg^2 \beta c^3 }
\label{eq:rhoMdot}
\end{equation}
where we have assumed a disk height of $r$ and an inflow velocity near
the horizon of $\beta c$.  The jet magnetic field will accumulate until it
balances the ram pressure of the accretion flow, and thus we arrive at
the first magnetic field strength estimate of
\begin{equation}
B \approx
\sqrt{8\pi \rho \beta^2 c^2}
\approx 
\left(\frac{2 L \beta}{\eta_d \Rg^2 c}\right)^{1/2}
\approx
3\times10^2 \eta_{d,-3}^{-1/2} L_{42}^{1/2} \beta^{1/2}~\G \,.
\label{eq:Brho}
\end{equation}

It is not clear that the mm-wavelength emission is associated with an
accretion flow.  At 7~mm the jet dominates the source morphology \citep{2008JPhCS.131a2053W}.
Thus, we present a second estimate based upon the assumption that the
jet is Poynting dominated near the black hole and relating the
mechanical power at small and large radii.  The Poynting luminosity
associated with black-hole driven jets is \citep{2010ApJ...711...50T}:
\begin{equation}
L_{\rm EM} = \frac{k}{4 \pi c} \Omega_{\mathrm{H}}^{2} \Phi^{2},
\label{eq:lem}
\end{equation}
where $k \approx 0.045$ for parabolic jet geometry, $\Phi$ is the poloidal
magnetic flux threading the black hole, and
$\Omega_H$ is the angular velocity of the horizon of the black hole,
which encodes the dependence upon the black hole spin.\footnote{Note
  that the low-spin approximation given by Equation~\eqref{eq:lem} remains
  accurate for spins $a\lesssim0.95$, beyond which it over-estimates
  the true power by up to $30$\%. For a higher order expansion
  that remains accurate for all spins, see
  \citet{2010ApJ...711...50T}. See also \citet{2012JPhCS.372a2040T}
  for the comparison of various approximation to jet power.}  

A variety of estimates of the total jet power in M87 have been
obtained across a wide range of scales.  Radio lobe measurements
probe emission extending over 30~kpc, and have resulted in estimates
of $L_{\rm jet}\approx10^{44}~\erg\,\s^{-1}$
\citep{deGasperin_etal:12}.  This is similar to
estimates arising from the modeling of the complex of knot features
on kpc-scales \citep{Owen-Eile-Kass:00}, and on 60~pc scales from
interpreting the HST-1 complex as a recollimation shock
\citep{Staw_etal:06}.  Thus, we adopt this estimate as a
conservative estimate of the total jet power.  For a characteristic
black hole spin, $a=0.9$, equating these then gives a poloidal
magnetic field strength estimate at the event horizon:
\begin{equation}
B_P = \frac{\Phi}{A} \approx 2.5\times10^2~\G,
\label{eq:BPest}
\end{equation}
where $A = 2\pi(a^2r_g^2+3r_{\rm H}^2)/3$ is the
area of the one of the two hemispheres of the event horizon, where
$r_{\rm H}=\Rg(1+\sqrt{1-a^2})$ is the radius of the black hole event horizon.

Despite the significantly different physics being invoked, the two
estimates are quite similar.  While this is not accidental (thick
disks are believed to be a necessary component of jet formation), it
does critically depend upon the low radiative efficiency generally
implicated in vastly sub-Eddington sources (such as M87).  

At the distances of the stagnation surface, which lies inside the
light cylinder, the magnetic field is
poloidally-dominated. 
If we now assume that the jet is
roughly parabolic, i.e., $R_j/\Rg\approx (r/\Rg)^{1/2}$, within the gap
region the above estimates imply
\begin{equation}
B\approx \frac{\Phi}{\pi R_j^2} \approx 35~\G\,.
\label{eq:Best}
\end{equation}
for a typical stagnation surface distance of $r\sim10\Rg$.

\subsubsection{Synchrotron Cooling Times}
In the jet frame, the observed synchrotron cooling timescale is
\begin{equation}
t'_{\rm sync} 
= 
\frac{3 m c^2}{4 \sigma_T u'_B c \gamma'}
=
\frac{6\pi m c^2}{\sigma_T {B'}^2 c} \sqrt{\frac{\nu_B'}{\nu'}}
=
\frac{3}{\sigma_T}
\left(\frac{ 2\pi e m c}{{B'}^3 \nu'}\right)^{1/2}\,,
\end{equation}
where $\nu'$ and $\nu_B'\equiv eB'/2\pi m_e c$ are the observation and
cyclotron frequencies {\em in the jet frame}, and
$\gamma'$ is the Lorentz factor of nonthermal particles, again measured
in the jet frame.  To relate this to emission in the lab frame, note
that $\nu' = \nu \Gamma\left(1-\beta\cos\Theta\right)$
and since the magnetic field is likely to be dominantly toroidal,
$B' \approx B/\Gamma$.  Hence, as measured in the lab frame,
\begin{equation}
\begin{aligned}
t_{\rm sync} 
=
t'_{\rm sync}\Gamma
&\approx 
\frac{3}{\sigma_T}\left[\frac{ 2\pi e m c \Gamma^4}{B^3\nu (1-\beta\cos\Theta)} \right]^{1/2}\\
&\approx
1.8\times10^5 B_{1.5}^{-3/2} \nu_{11.5}^{-1/2} \,\s\,,
\end{aligned}
\label{eq:tsync}
\end{equation}
where $B=30 B_{1.5}\,\G$ and $\nu=300\nu_{11.5}\,\GHz$ and the
remainder of the quantities are given by our fiducial M87 values.

Evident in Equation~\eqref{eq:tsync} is the strong dependence of
$t_{\rm sync}$ on $B$ and $\Gamma$, and therefore position.  As the
jet propagates $B$ decreases and $\Gamma$ increases in a fashion that
depends on the specific jet structure, resulting in a corresponding
rapid increase in $t_{\rm sync}$.  Generally, magnetic flux
conservation implies that the poloidal and toroidal components of the
magnetic field within the jet scale $\propto R^{-2}$ and 
$\propto R^{-1}$, respectively (it is for this reason that despite having
similar strengths near the black hole, jet magnetic fields quickly
become strongly toroidally dominated).  For a parabolic magnetic
field, $R\propto z^{1/2}$, which we assumed when obtaining
Equation~\eqref{eq:tsync}.  Similarly, $\Gamma$ is expected to rise
$\propto R$.  However, in practice the Lorentz factor of the jet in
M87 appears to asymptote to $\approx5$ at large radii, limiting its
impact.  Regardless, near the stagnation surface these imply that
$t_{\rm sync}$ will initially grow $\propto z^{7/4}$, eventually
slowing to $\propto z^{3/4}$.

The implication for the image structure is obtained by comparing
$t_{\rm sync}$ to the outflow timescale,
\begin{equation}
t_{\rm outflow} = \frac{z}{\beta c} 
\approx
3.3\times10^5 \left(\frac{r}{10\Rg}\right)~\s\,,
\label{eq:toutflow}
\end{equation}
which grows linearly with jet height.
When $t_{\rm sync}>t_{\rm app}$ the relativistic leptons cannot
efficiently cool, and therefore the nonthermal particle density is
effectively conserved.  At moderate distances from jet base
($z\gtrsim15\Rg$) the above estimates suggest that this is the
case. At much larger distances where the Lorentz factor saturates
($z\gtrsim10^3\Rg$) the synchrotron cooling becomes efficient.  

Near the stagnation surface $t_{\rm sync}$ is comparable to 
$t_{\rm outflow}$.  However, this is complicated by the large initial
bulk Lorentz factors resulting from the acceleration across the gap.
As a result, a substantial fraction of leptons still resides in the
ordered non-thermal relativistic component that emerged from the polar
cascade propagating along magnetic field lines and therefore
geometrically immune to synchrotron cooling.  The study of how 
these nonthermal electrons isotropize and cool is beyond the scope of
this paper. Thus, for simplicity we will ignore synchrotron cooling in
the estimation of the energetic lepton population of the jet.

\subsubsection{Seed Photon Density Estimates}
\label{sec:seed-photon-density-m87}
The source of the soft seed photons in M87 is not immediately clear
due to potential contamination from the jet itself.  Nonetheless, it
is now certain that the millimeter-wavelength emission is associated
with compact features with scales comparable to that of the horizon
\citep{2012Sci...338..355D}.
This provides circumstantial evidence that the observed millimeter,
infrared, and optical emission arise very near the black hole itself.
Thus, even if the emission is from within the jet itself, at the jet
heights of relevance the bulk Lorentz factors are expected to be small
(of order unity), and hence not strongly beamed.  As a consequence, we
estimate of the seed photon density from the observed emission directly.

\begin{figure}[h!]
\begin{center}
\includegraphics[width=\columnwidth]{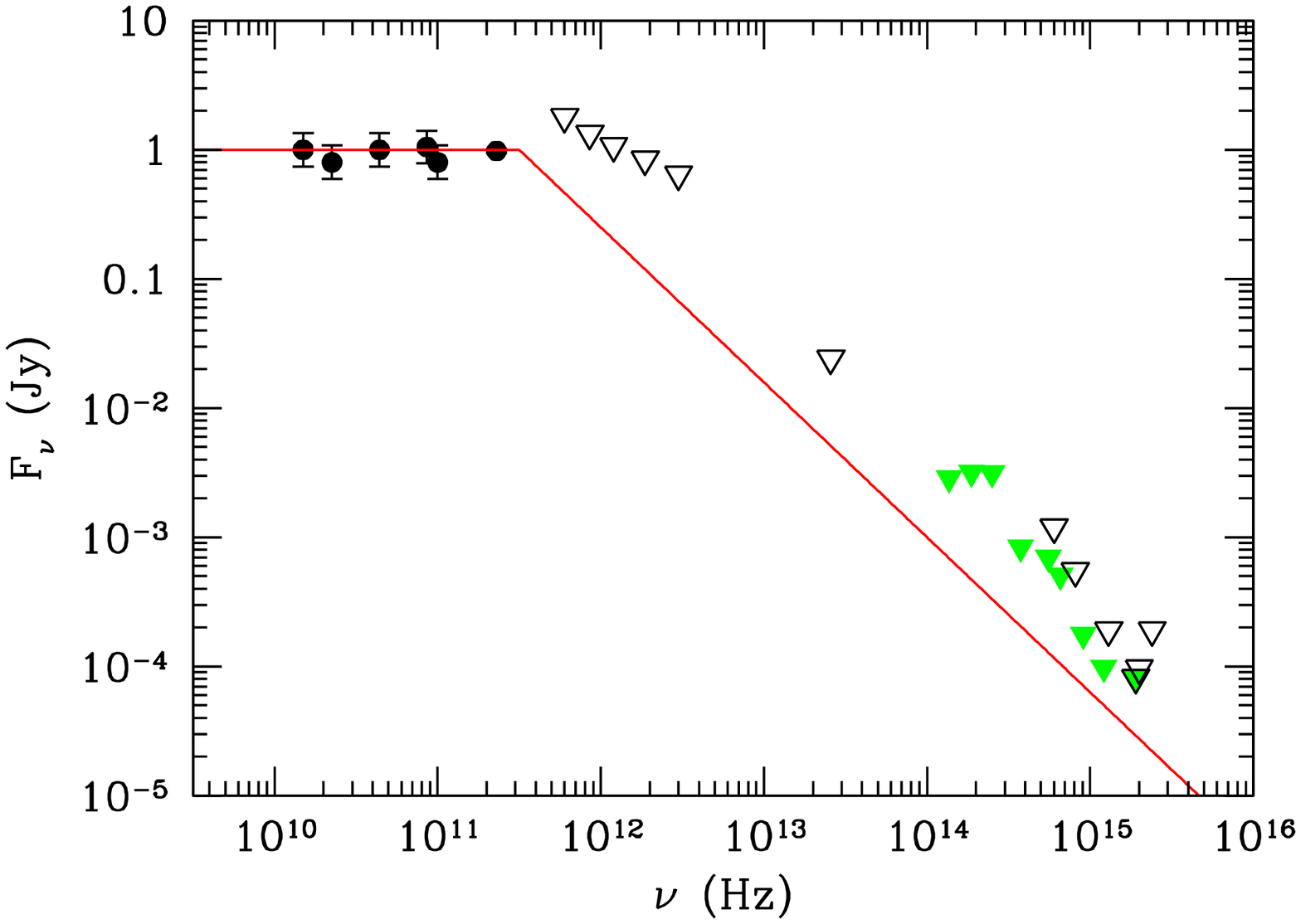}
\end{center}
\caption{Spectral energy distribution for M87 from the
  radio to the optical, constructed from the literature
  \citep[additional details will be published elsewhere]{Spen-Juno:86,Baat_etal:92,Juno-Bire:95,Spar-Bire-Macc:96,Lons-Doel-Phil:98,Bire-Juno-Livi:02,Ly-Walk-Wrob:04,Kova_etal:05,List-Homa:05,Petr_etal:07,Perl_etal:07,Lee_etal:08,Baes_etal:10,2012Sci...338..355D}.  
  Green filled triangles show a subset of the infrared-optical flux limits collected in
  \citet{2015ApJ...805..179B}.
  Filled circles and open triangles correspond
  to flux measurements in which the source is resolved and unresolved,
  respectively; we take the filled circles as directly
  indicative of the near-horizon seed photon distribution and the open
  triangles as upper limits.  Errorbars indicate variability, not
  intrinsic measurement uncertainty.  For comparison, a simple broken
  power-law with high-frequency spectral index $\alpha=1.2$, as
  employed in the text, is shown for reference.}\label{fig:M87SED}
\end{figure}

The SED of M87, shown in Figure \ref{fig:M87SED}, is reasonably well
fit by a broken power law, i.e., $F_\nu\propto\nu^{-\alpha}$ where
$\alpha$ evolves rapidly as the source becomes optically thin (see
  Fig.~\ref{fig:M87SED}):
\begin{equation}
\alpha\approx
\begin{cases}
0 & \nu<300~\GHz\\
1.2 & \nu\ge300~\GHz\,.
\end{cases}
\end{equation}
The flux density at the break is roughly 1~Jy.  If all of the emission
above 300~GHz arises in a radiatively inefficient accretion disk, then
the soft photon energy density is roughly
\begin{equation}
u_s \approx 0.01~\erg~\cm^{-3}\,,
\label{eq:usm87}
\end{equation}
where we have now fixed the location of the stagnation surface to
$10\Rg$.  This is well above the densities necessary for inverse
Compton losses to dominate the limits upon the acceleration of the
leptons within the gap. 

\subsubsection{Implied Nonthermal Particle Properties}
If the sub-mm luminosity arises due to optically thin emission of the
jet launching region, as suggested by the recent mm-VLBI observations \citep{2012Sci...338..355D},
it implies a rather soft nonthermal particle density, with
$dn/d\gamma\approx\gamma^{-3.4}$.  The source size is roughly
$40~\muas$, implying that the associated millimeter flux is related to
the particle density by
\begin{equation}
\begin{aligned}
F_{\mm}
&\approx
\frac{\sqrt{27}e^2\nu_B}{8\pi^2 c}
\left(\frac{3\nu_B}{\nu}\right)^{\alpha}
\frac{2\alpha \gamma_{\rm min}^{2\alpha}}{\alpha+1} 
\Gamma\left(\frac{\alpha}{2}+\frac{11}{6}\right)
\Gamma\left(\frac{\alpha}{2}+\frac{1}{6}\right)
\Omega r n\\
&=
4.9\times10^{-2}
\left(\frac{\theta}{40\muas}\right)^2
\left(\frac{r}{10\Rg}\right) n B_{1.5}^{1+\alpha}
~\Jy\,,
\end{aligned}
\label{eq:Fmm}
\end{equation}
where $\nu_B\equiv eB/2\pi m_e c$ is the cyclotron frequency,
$\Gamma(x)$ is the standard gamma-function, $\nu=300$~GHz, and we have
assumed a typical lower cutoff on the accelerated particle
distribution of $\gamma_{\rm min}\approx10^2$.  Comparing this to an
observed flux of roughly 1~Jy and assuming the magnetic field estimate
in Equation (\ref{eq:Best}) gives an approximate nonthermal particle  
density estimate of
\begin{equation}
n\approx 15~\cm^{-3}\,,
\end{equation}
consistent with efforts to quantitatively model the emission region of
M87 on horizon scales (Broderick et al., in preparation).

\subsection{Particle Acceleration at the Stagnation Surface}

We can now estimate the properties of the gap, and the associated
accelerated particles, from M87's parameters given above. This gives
our fiducial model for M87. The asymptotic lepton Lorentz factor is
determined by inverse Compton losses and is roughly 
\begin{equation}
\gamma_{\rm max} \equiv \gamma_{\rm max, IC}
\approx
1.6\times10^9\,,
\label{eq:gammamaxm87}
\end{equation}
and this sets our fiducial value of electron energy,
$E_0$. Whereas its numerical value is known only to a factor of few, we
choose not to round this value in order to get a self-consistent answer.
This corresponds to a seed photon pair-production threshold of
\begin{equation}
\epsilon_{\rm th}
\approx
1.2~\meV
\quad\Rightarrow\quad
\lambda_{\rm th}
\approx
1.1~\mm\,.
\label{eq:ethm87}
\end{equation}
Therefore, photons at and just below the spectral break in M87's SED
will contribute to pair production within the gap.

The length scale over which the asymptotic lepton Lorentz factors are
obtained is (see Equation~\ref{eq:ellIC})
\begin{equation}
\ell_{\rm IC} \approx 5.5\times10^{10}~\cm = 5.6\times10^{-5}\Rg\,,
\label{eq:ellicm87}
\end{equation}
and the typical mean free path of the up-scattered photons to pair
production is (see Equation~\ref{eq:ellgg})
\begin{equation}
\ell_{\gamma\gamma}
\approx
1.7\times10^{12}~\cm
\approx
1.7\times10^{-3}\Rg\,.
\label{eq:ellggm87}
\end{equation}
Hence, as anticipated, both the acceleration and subsequent pair
production within the gap occurs on scales much smaller than the
typical gap thickness that might be expected on global considerations
(i.e., $\Rg$).  The resulting gap thickness is then roughly (see 
Equation~\ref{eq:gapthickness})
\begin{equation}
\Delta \approx 4.3\times10^{11}~\cm\,,
\label{eq:deltam87}
\end{equation}
implying a corresponding density at the top of the gap of (see Equation~\ref{eq:ngap})
\begin{equation}
n_{\rm g} \approx 2.2\times10^{-2}~\cm^{-3}\,.
\label{eq:ngm87}
\end{equation}
We note that since $\Delta$ by about an order of magnitude exceeds
$\ell_{\rm IC}$, over which the pairs attain their terminal Lorentz
factor, our earlier assumption that the pairs instantaneously attain
their terminal Lorentz factor is justified.

\subsection{Post-Gap Cascade}
 
\label{sec:post-gap-cascade}
Given the threshold seed photon energy of 1.2~meV, the minimum
gamma-ray energy for which the seed photons bath is optically-thick to
pair production is 
$\epsilon_{\gamma,\rm min} = 640~\GeV$, hence the
asymptotic Lorentz factor for M87 is (see Equation~\ref{eq:ggg})
\begin{equation}
\gamma_{\gamma\gamma} = 2.6\times10^6\,,
\end{equation}
implying enhancement in the number density due to the post-gap cascade
of roughly $n_\infty/n_g = 670$ (see Equation~\ref{eq:enhancement}).  

\begin{figure*}
\begin{center}
\includegraphics[width=1.0\textwidth]{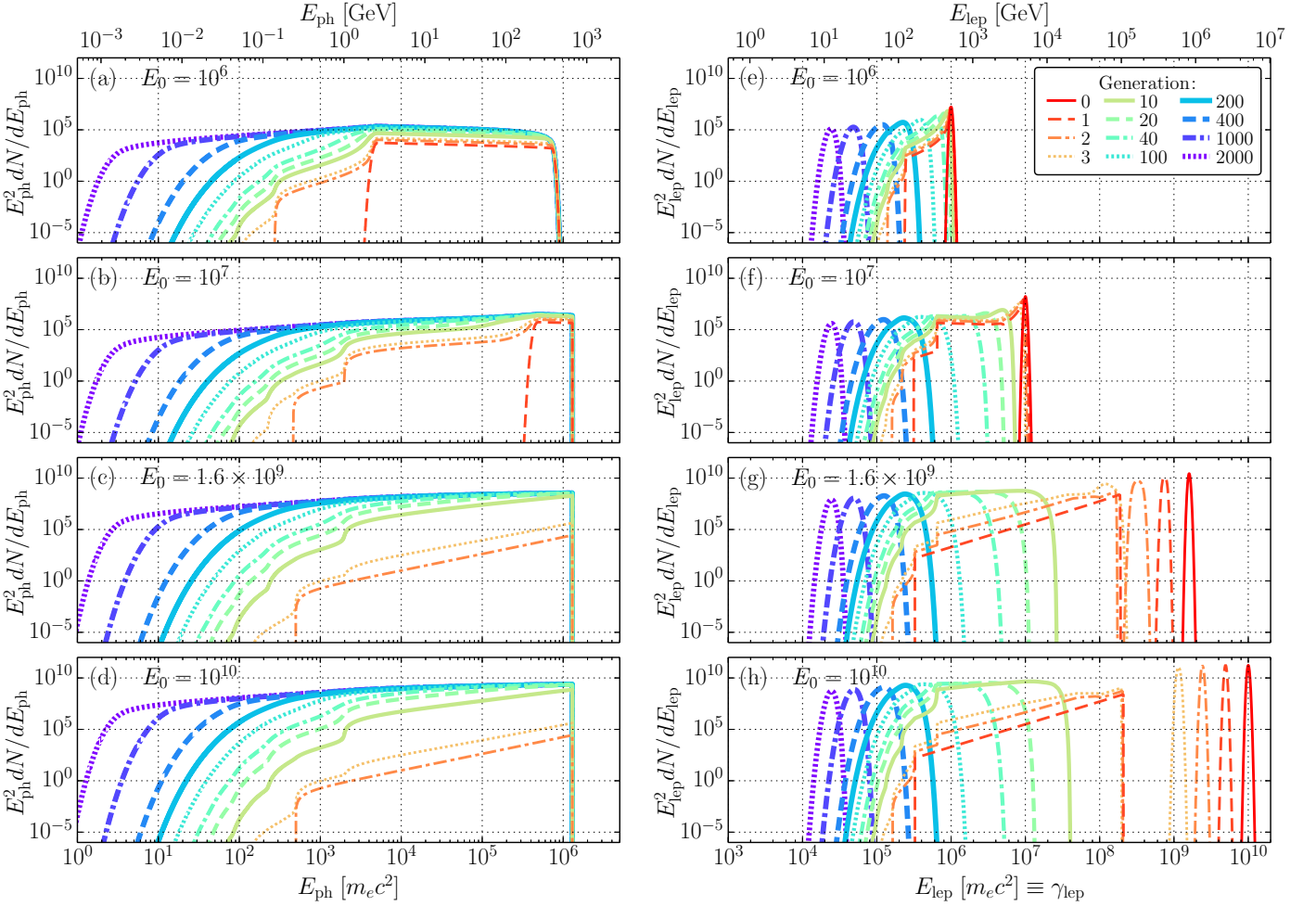}
\end{center}
\caption{A sequence of the resulting energy distributions for escaping
  photons (left column) and leptons (right column) for different
  values of initial energy of the electron, shown in different rows,
  from top to bottom: $E_0/m_e c^2 = 10^6, 10^7, 1.6\times10^9,
  10^{10}$. Different generations of lepton and photon energy spectra
  are shown with lines of different colors, thicknesses and line types (see the legend). We
  start at generation zero with a single electron at 
  energy $E_0$. The electron up-scatters a seed photon into gamma rays
  and cools into a power-law energy distribution. The gamma ray
  pair-produces off of another seed photon (see main text for
  details). Gamma rays, which are not energetic enough to
  pair-produce, escape, and their spectrum is shown in the left
  column. Note that at generations of $\approx100$ and greater, the shape
  of the spectrum of leptons and photons is essentially independent of
  the initial energy of the electron, $E_0$: it is only the
  normalization of the spectra that depends on $E_0$, not their shape.
}\label{fig:dnde}
\end{figure*}
To verify this analytical estimate, we carried out numerical
integration of the pair cascade equations
\eqref{eq:lep}--\eqref{eq:ph}, as described in Section~\ref{sec:NAE}. 
We choose the following parameters to describe the seed photon
spectrum in our fiducial model of M87: 
$\alpha = 1.2$, $\epsilon_m = 1.2$~meV, $\epsilon_M = 0.8$~eV.
We start with an initial distribution that corresponds to
an electron emerging from the top of the gap at an energy $E_0$.  For
numerical convenience, we represent its energy distribution with a
narrow Gaussian distribution shown in Figure~\ref{fig:dnde}(e)-(h)
with thin solid dark red line.  Different rows show distributions for 
different values of $E_0/m_e c^2$:
$10^6$, $10^7$, $1.6\times10^9$ (fiducial value), and $10^{10}$. 
We then evolve the system of equations \eqref{eq:lep}--\eqref{eq:ph}
numerically, as described in Section~\ref{sec:NAE}, and show subsequent distributions in
Figure~\ref{fig:dnde} with colored lines (see the
legend). 

We adopt as our fiducial model pair cascade for M87 the case when 
$E_0 = 1.6\times 10^{9}m_ec^2$, shown in Figure \ref{fig:dnde}(g).
The electron cools
(see Equation~\ref{eq:lepic}) due to the upscattering of a seed photon
into a $\gamma$-ray photon, which in turn pair produces on an
additional seed photon (see Equation~\ref{eq:lepgg}).  These two
contributions to lepton energy distribution are clearly visible in
Figure~\ref{fig:dnde}(h): after the first scattering, i.e., in generation 1,
the IC-cooled electron distribution is a power-law extending from
$\gamma_{\rm lep}\approx0.5\epsilon_{\gamma,\rm min} = 6\times10^5$ to
$\gamma_{\rm lep}=2\times10^8$ and the new pairs form a Gaussian
distribution centered around $\gamma_{\rm lep}\approx7\times10^8$. Note that
the lower-energy cutoff in the power-law emerges because the seed
photon bath is transparent to gamma-rays with energy less than
$\epsilon_{\gamma,\rm min}$.

\begin{figure}
\begin{center}
\includegraphics[width=\columnwidth]{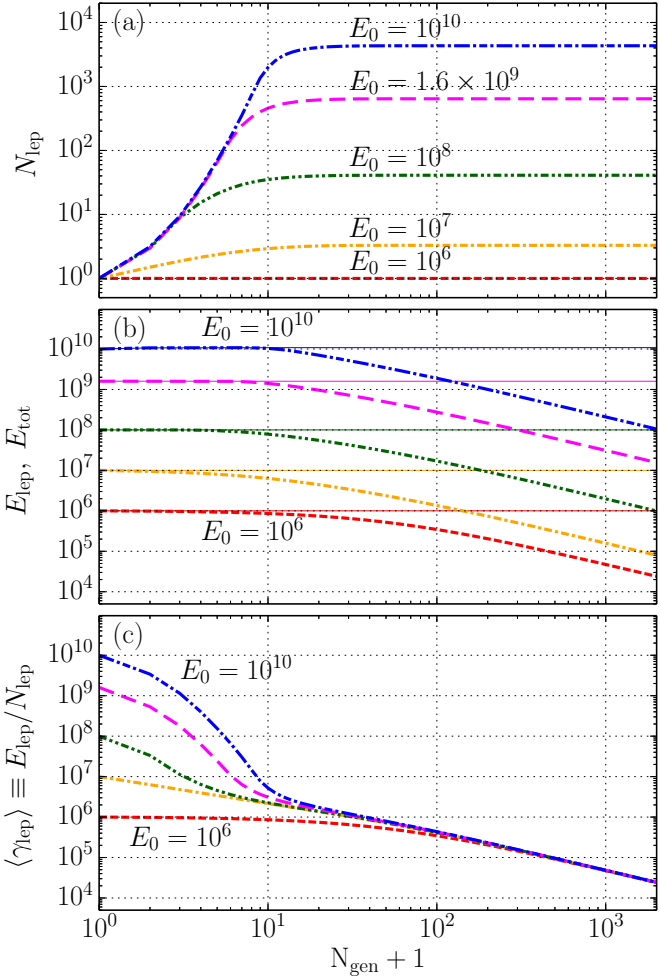}
\end{center}
\caption{Evolution of various quantities in pair cascade with
  generation for different values of initial lepton energy, $E_0$,
  which is measured in units of $m_e c^2$.
  Panel (a) shows the dependence on the generation number of density
  enhancement, $N_{\rm lep}\equiv n_\infty/n_g$. Panel (b) shows the
  dependence of the total energy in leptons and photons with colored
  solid lines and the lepton total energy, $E_{\rm lep}$, with other
  lines, for different values of initial lepton energy, $E_0$. Panel
  (c) shows the evolution with the generation number of the average
  lepton Lorentz factor, $\langle\gamma_{\rm lep}\rangle$.
}\label{fig:cascade}
\end{figure}

Numerically-obtained values of density enhancement vs. generation
number are shown in Figure~\ref{fig:cascade}(a), for different values
of $E_0$. It is clear that the density enhancement saturates around
generation $N_{\rm gen}\gtrsim{\rm few}\times10$.  The numerical
result for density enhancement in our fiducial model of M87,
$n_\infty/n_g = 640$, is shown with long-dashed magenta line and is in
excellent agreement with the analytic expectation, discussed above and
given by Equation~\eqref{eq:enhancement}, $n_\infty/n_g\approx670$.
This yields a jet lepton density near the stagnation surface of
\begin{equation}
n_\infty \approx 15~\cm^{-3}\,.
\end{equation}
In the absence of nonlinear plasma phenomena, these energic pairs
will inverse-Compton cool over roughly a single jet scale height,
i.e., within a distance $r$, at which point they will have undergone
large-angle deflections.  These leptons, now misaligned with the
background magnetic field, will subsequently synchrotron cool and
isotropize.  The rate of this cooling depends on the microscopic
Lorentz factors and becomes comparable to the outflow timescale only
when $\gamma\approx10^{2-3}$ on scales of the gap height.  Thus the
resulting population is in excellent agreement with that required to
produce the observed sub-mm emission.

As is also clear from Figure~\ref{fig:cascade}(a), higher initial
energy leads to a higher density enhancement. Similarly,
Figure~\ref{fig:cascade}(b) shows that the total energy and the energy in
leptons are higher for higher $E_0$. 
The flatness of the solid lines in Figure~\ref{fig:cascade}(b)
demonstrates that the total energy of the cascade (carried by leptons
and photons) is conserved at high accuracy by our numerical scheme.
Note that as seen in Figure~\ref{fig:dnde}, there exist asymptotic
properties of the pair cascade that are independent of $E_0$.  For
example, it is explicitly apparent in Figure~\ref{fig:cascade}(c) that
the average lepton Lorentz factor, $\langle\gamma_{\rm lep}\rangle$,
is independent of $E_0$ for $N_{\rm gen}\gtrsim100$ and is controlled
completely by $\tau_{\gamma\gamma}$.

\begin{figure}
\begin{center}
\includegraphics[width=0.49\columnwidth]{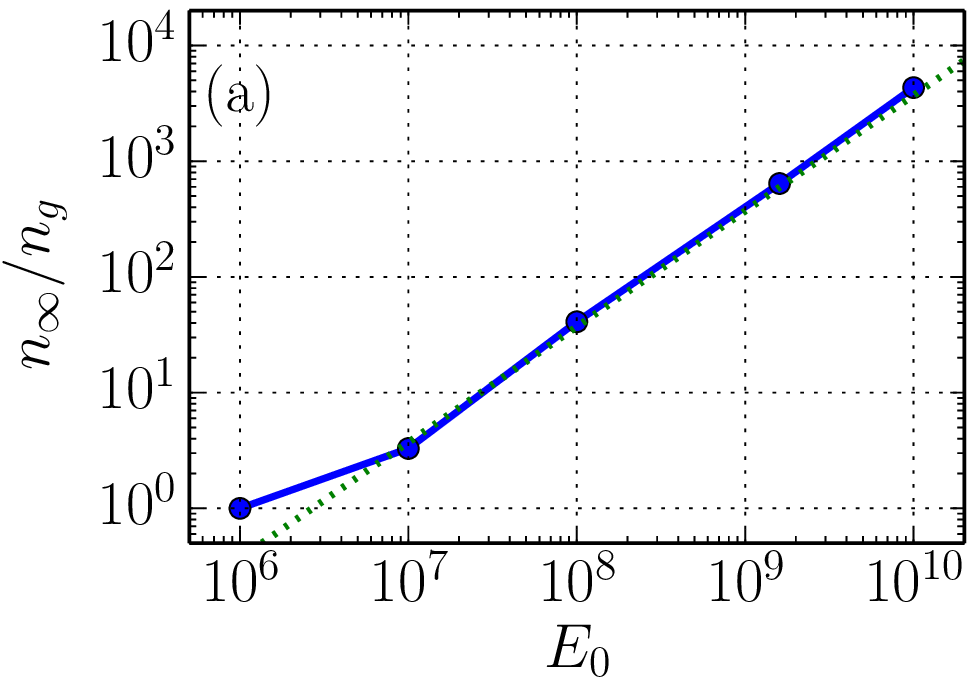}\
\includegraphics[width=0.49\columnwidth]{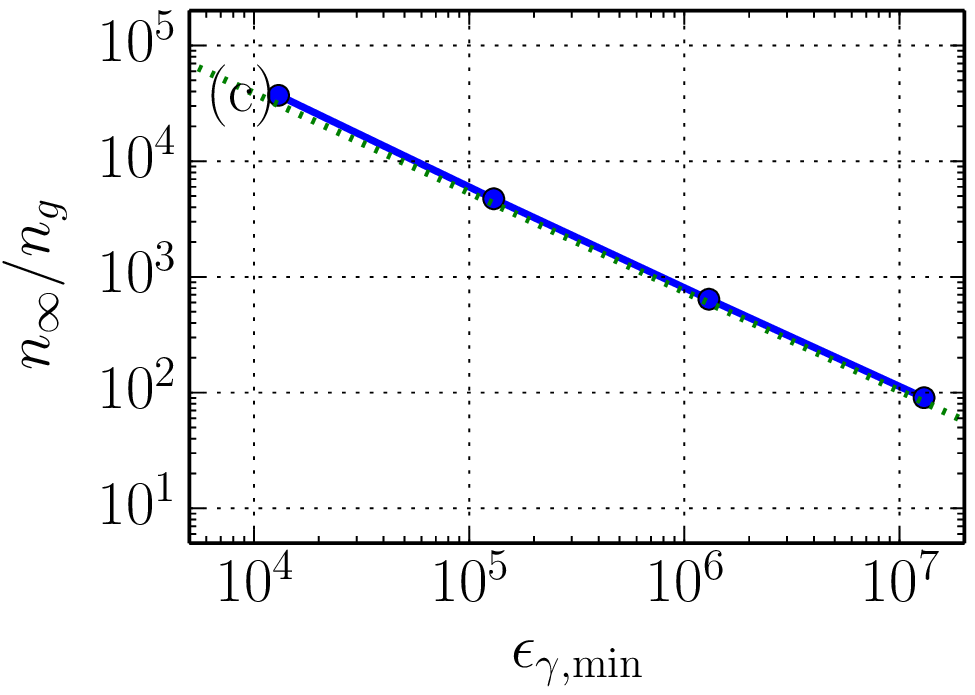}\\
\includegraphics[width=0.49\columnwidth]{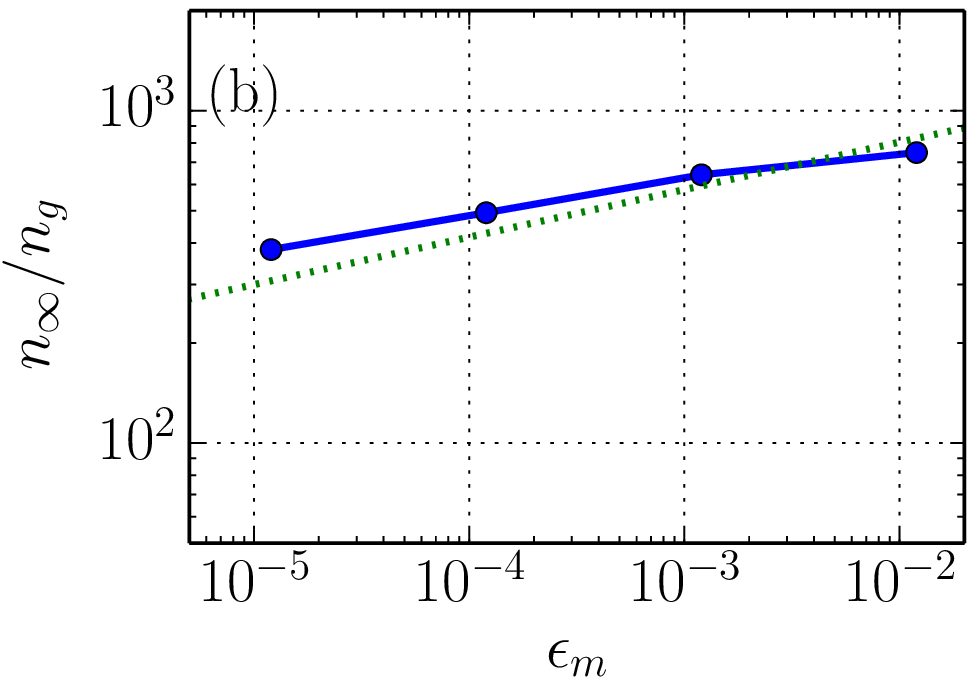}\
\includegraphics[width=0.49\columnwidth]{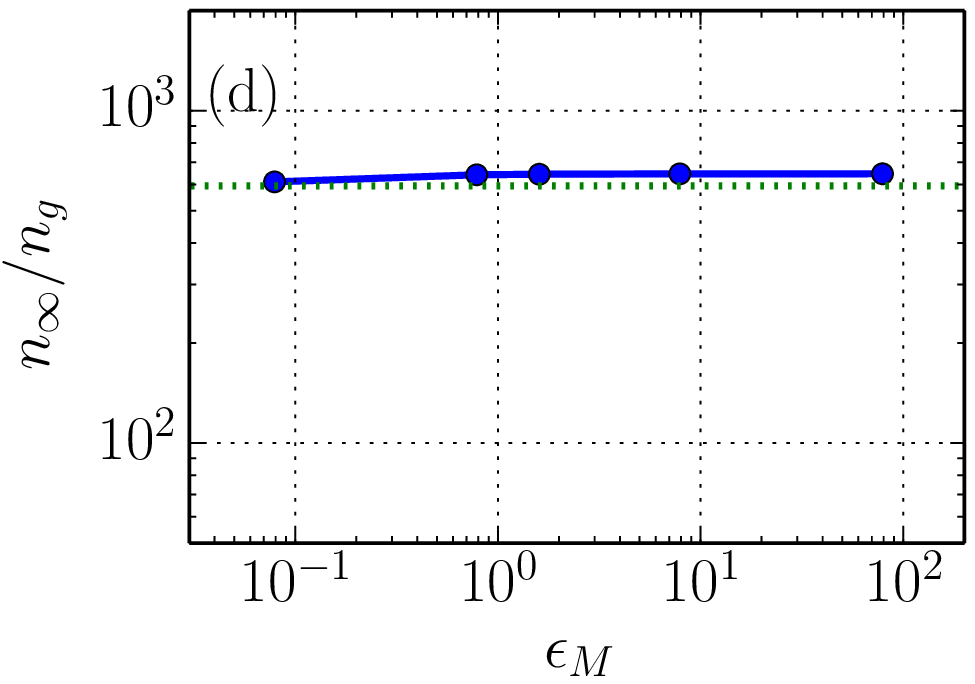}\\
\end{center}
\caption{Dependence of density enhancement factor, $n_\infty/n_g$, 
on the parameters of the cascade. Connected
  blue dots show simulation results and the green dotted lines
  show the analytical expression given by
  Equation~\eqref{eq:enhancement}. Panel~(a) shows the dependence of
  density enhancement factor, $n_\infty/n_g$ on the initial energy 
  $E_0$ of the lepton, panel~(b)
  on the lower energy cutoff in the seed photon distribution
  $\epsilon_m$, panel~(c) on the minimum energy of gamma-rays able to
  pair produce $\epsilon_{\gamma,\rm min}$ (see Equation~\ref{eq:glim}), and
  panel~(d) on the upper energy cutoff in the seed photon
  distribution. The agreement between the model and the simulation is
  better than $30$\%.}
\label{fig:nvse0}
\end{figure}

We investigated the dependence of density enhancement factor on key
parameters of our model, and the results are shown in
Figure~\ref{fig:nvse0}. The blue dots connected by solid lines show the
numerical results; they are in excellent agreement with the analytical
scaling given by
Equation~\eqref{eq:enhancement}. The
density enhancement is most sensitive to the initial electron energy,
$E_0$: $N_{\rm lep,\infty}\propto E_0$ [see Figure~\ref{fig:nvse0}(a)], and the gamma-ray energy below
which the seed photon bath is optically-thin to pair production,
$\epsilon_{\gamma,\rm min}$: 
$N_{\rm lep,\infty}\propto\epsilon_{\gamma,\rm min}^{\alpha/(2\alpha-1)}\approx \epsilon_{\gamma,\rm min}^{0.86}$ 
(see Figure~\ref{fig:nvse0}(c)). However, if we now recall that both 
$E_0 = \gamma_{\rm max,IC}m_e c^2$ and 
$\epsilon_{\gamma,\rm min}\propto \gamma_{\rm max,IC}$ depend on
$\gamma_{\rm max,IC}$ (see Equation~\eqref{eq:glim}), we find that
these dependencies nearly cancel out, and that density enhancement is
most sensitive to the density of seed photons $n_s$, or source
luminosity:
\begin{equation}
\frac{n_\infty}{n_g}
\propto
\left(\gamma_{\rm max,IC}\epsilon_b\right)^{(\alpha-1)/(2\alpha-1)} n_s^{1/(2\alpha-1)}
\approx
\left(\gamma_{\rm max,IC}\epsilon_b\right)^{0.14} n_s^{0.71}\,. 
\label{eq:Nlepinfty}
\end{equation}
Importantly, the numerical results are insensitive to the upper energy
cutoff of the seed photon spectrum, as shown in Figure~\ref{fig:nvse0}(d).

\begin{figure}
\begin{center}
\includegraphics[width=0.9\columnwidth]{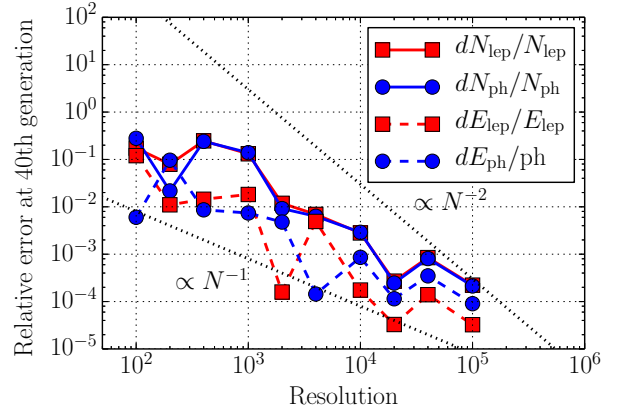}
\end{center}
\caption{Dependence of fractional error on resolution in our fiducial
  model  (evaluated at 40th lepton generation) for various quantities:  
  lepton number, $N_{\rm lep}$, emitted photon number, $N_{\rm ph}$,
  lepton energy, $E_{\rm lep}$, emitted photon energy, $E_{\rm ph}$. 
  The error is computed relative to a high-resolution model with
  $2\times10^5$ spectral energy bins, and exhibits 
  linear convergence. 
  At our fiducial resolution of
  $10^4$ energy bins, the relative error is less than $0.3$ per cent
  in all quantities.} 
\label{fig:convergence}
\end{figure}

We ensured that our simulation results are numerically converged.  For
this, we varied the number of energy bins for more than $3$ orders of
magnitude and computed the error relative to the highest-resolution
simulation carried out using $2\times10^5$ energy bins.
As seen in
Figure~\ref{fig:convergence}, the simulation results converge at
first order with increasing energy resolution. 
For the simulations described here, we employed a fiducial
resolution of $10^4$ energy bins, for which the relative error is less
than $0.3$ per cent in all quantities.

The asymptotic Lorentz factor after the pair cascade saturates, set by
when the inverse-Compton cooling length is comparable to the jet scale
height, is roughly $10^4$.  This implies direct synchrotron
emission up to $\approx10$--100~eV, well into the ultraviolet.

\subsection{Direct Inverse Compton Signal}
The direct inverse Compton signal from the stagnation surface itself
should be visible up to $\epsilon_{\rm VHEGR}\approx640$~GeV, above which
it will be absorbed by the pair cascade.  As is clear from
Figure~\ref{fig:dnde}, the direct IC signal from the cascade has a flat
energy spectrum in $E^2 dN/dE$, with every decade in energy carrying
approximately the same amount of energy. This is consistent with the
analytical expectation given in Appendix~\ref{sec:DICFDE}.
Thus, it is possible that the observed very-high energy gamma-ray
emission is associated with the stagnation surface 
\citep[see, e.g.,][]{2011ApJ...730..123L}.  Emission in the band pass
of \emph{Fermi}/LAT, around $\epsilon_\gamma \approx 10$~GeV, is
mostly due to Compton up-scattering of seed photons near the peak of
the SED, $\epsilon_s \approx1.2$~meV, by electrons with Lorentz
factors
$\gamma\approx(\epsilon_\gamma/2\epsilon_s)^{1/2}=2\times10^6$, 
which is comparable to $\gamma_{\gamma\gamma}$.
The associated power per lepton is 
\begin{equation}
P_{\rm 10~GeV}\approx4\sigma_T u_s c \gamma^2/3.
\label{eq:p10gev}
\end{equation}
The highly relativistic nature of the lepton distribution produces a
large relativistic aberration, beaming the up-scattered gamma rays
within a cone with an opening angle $1/\gamma$ around the original
electron momentum vector, and thus concentrating the emission in a solid
angle of $\Omega_\gamma=\pi/\gamma^2$.  However, it also limits the
region of the stagnation surface that can be viewed to an area
$\Omega_\gamma r^2$.  Finally, inverse-Compton cooling also sets a
length scale over which we expect to find particles that can produce
sufficiently energetic up-scattered gamma rays,
$\ell_{\rm IC, 10~GeV}$ typically much smaller than $\Rg$, and here
roughly $0.05\Rg$ (see Equation~\ref{eq:ellIC}).  Thus, the typical
anticipated 10 GeV flux produced near the stagnation surface is
independent of the beaming, assuming that the viewing angle is within
the jet opening angle:
\begin{equation}
\begin{aligned}
\left.F_\epsilon\right|_{\rm 10~GeV} 
&\approx
\frac{P_{\rm 10~Gev}}{\Omega_\gamma}
n_\infty
\frac{\Omega_\gamma R_j^2}{D^2}
\ell_{\rm IC, 10~GeV}
=
\gamma m c^3 n_\infty \frac{\pi R_j^2}{D^2}
\\
&=
7.1\times10^{-9} ~\erg~\cm^{-2}~\s^{-1}\,.
\label{eq:Fe10Gev}
\end{aligned}
\end{equation}
In principle, all leptons with Lorentz factors above $\gamma$ will
contribute to the energy flux.   In practice, this increases the flux
near 10~GeV by a factor of less than two (see Appendix
\ref{sec:DICFDE}), to 
\begin{equation}
\begin{aligned}
\left. \epsilon F_\epsilon\right|_{\rm 10~GeV}
&\approx 
\frac{12(\alpha-1)}{2\alpha-1} \gamma m c^3 n_\infty \frac{\pi R_j^2}{D^2}\\
&= 1.2\times10^{-8}~\erg~\cm^{-2}~\s^{-1}\,.
\end{aligned}
\end{equation}  
This is considerably larger than the
$(7.5\pm5.9)\times10^{-12}~\erg~\cm^{-2}~\s^{-1}$ reported for M87 by
\emph{Fermi} in the first \emph{Fermi}/LAT catalog of $>10$~GeV
sources \citep[1FHL,][]{1FHL}.
It is, however, consistent with the distance-adjusted fluxes of other
gamma-ray bright blazars in the \emph{Fermi} sample \citep[see, e.g.,][]{2LAC}.

The disparity is almost certainly due to the high degree of beaming
anticipated.  At 10~GeV, the high energy emission is beamed within
0.1'' of the original electron direction.  Beyond this angle, the energy
of the up-scattered photons drops dramatically; at $10^\circ$ the
typical up-scattered photon energy is only roughly 70 times larger
than that of the original seed photon, well below the energies of
interest.  Thus, the gamma-ray emission from the stagnation surface
would be expected to drop precipitously outside of the jet
half-opening angle.  This is a strong function of height, scaling
$\propto z^{-1/2}$ near the black hole, though at our fiducial height
of the stagnation surface this is $\theta_j\equiv R/z \approx18^\circ$.
Because the jet is collimating the tangent to the field
lines, which correspond to the range of angles over which the
relativistic leptons are directed and subsequent gamma rays emitted,
makes a considerably smaller angle with the jet axis, $\theta_b\approx0.5\theta_j=9^\circ$ for a
parabolic jet.  Both of these are well within the inclination implied
by radio observations of $25^\circ$, suggesting that it in this case
it would be rare to find substantial gamma-ray emission in M87.

\subsection{Total power of the cascade}
\label{sec:total-power-cascade}
Using Equation~\eqref{eq:Llep}, we obtain the total power dissipated by the
cascade,
\begin{equation}
  L_{\rm cascade} \approx 10^{43}\ {\rm erg\,s^{-1}}\,,
\label{eq:lcascadem87}
\end{equation}
corresponding to roughly 10\% of the total jet power,
$L_{\rm jet} \approx 10^{44}$~${\rm erg} {\rm s}^{-1}$.  Thus, most of
the jet power flows out in the form of Poynting flux and only 10\% of
it gets converted into particle energy at the stagnation surface and
the cascade.  Most of this power is radiated in the form of gamma-rays
and is not observable due to beaming away from our line of sight, as
we discussed above.  A small fraction of this power,
\begin{equation}
\label{eq:ptotcascade}
\begin{aligned}
  P_\infty 
  &= 
  \frac{\gamma_\infty n_\infty}{\gamma_{\rm max,IC} n_g} L_{\rm cascade}
  =
  \frac{\gamma_\infty}{\gamma_{\gamma\gamma}} L_{\rm cascade}\\
  &\approx 4\times 10^{-3} L_{\rm cascade}
  = 5\times 10^{40}\ {\rm erg\,s^{-1}},
\end{aligned}
\end{equation}
is carried asymptotically by the cascade and will be radiated via
synchrotron emission.  
This is roughly a factor of 20 smaller than the isotropic-equivalent
luminosity, consistent with the beaming correction inferred from
detailed source modeling \citet{2009ApJ...697.1164B}.\footnote{Note
  that since the angular size of the source is fixed in the estimates
  of the seed photon densities, the beaming correction does not affect
  $u_s$.}  This is not surprising since the asymptotic values of
density and Lorentz factor produced by the cascade, $n_\infty$ and
$\gamma_\infty$, are consistent with those necessary to explain M87's
radio emission, as discussed in \S\ref{sec:discussion}.

We note that in addition to total energy budget
our model has the potential to address an important morphological
feature of M87 jet that is pronounced in radio images at a frequency
of $43$~GHz \citep{2008JPhCS.131a2053W}: edge-brightening of the jet.
The electric field in the gap, given by Equation~\eqref{eq:ecorotation}, increases away
from the rotational axis.  This leads to density enhancement peaked
toward the edge of the jet (see Equations~\ref{eq:gammamaxIC} and
\ref{eq:Nlepinfty}) and therefore has the potential to lead to
stronger jet emission near the jet edges.  More detailed study of this issue
is warranted.

\section{Application to Sgr~A*}
\label{sec:application-sgr-a}

A second obvious application is to the putative jet in Sgr~A*, the
black hole at the center of the Milky Way.  Unlike M87, there is no
obvious radio jet in Sgr~A*, making the task of constraining its
properties somewhat less well defined.  However, there is some recent
evidence for a collimated outflow containing a non-relativistic
population of leptons \citep{Li-Morr-Baga:13}.  In this case the
putative jet lights up on parsec scales as a result of intersecting,
and shock heating upon, a stream of gas in Sgr~A*'s vicinity.  Again,
we will first review some relevant parameters for Sgr~A* and then
assess the implications of the particle acceleration model presented
in Section \ref{sec:pcass}.

This is complicated by the current uncertainty regarding the
dynamics and morphology of the region responsible for the observed
sub-mm emission.  This uncertainty manifests itself primarily in the
assumed magnetic field strength.  In sections
\ref{sec:SgrASP}--\ref{sec:SgrAPGC} we will assess in detail the
implications for a putative jet assuming that the observed emission
arises primarily in a RIAF.  In section \ref{sec:SgrAJet} we
consider the possibility of a self-consistent solution in which the
jet itself is responsible for the sub-mm emission.

\subsection{Sgr~A* Source Parameters} \label{sec:SgrASP}
The mass of and distance to Sgr~A* are the best known of any black
hole candidate, obtained from observations of the orbits of individual
stars, yielding $M=4.3\pm0.4\times10^6\,M_\odot$ and $8.3\pm0.4$~kpc,
respectively \citep{2008ApJ...689.1044G,2009ApJ...692.1075G}.  
As with M87, the SED peaks near 1~mm,
where the typical observed flux is roughly 3~Jy.  However, the
isotropic-equivalent luminosity is much lower, roughly 
$L\approx10^{36}~\erg~\s^{-1}\approx 10^{-9} L_{\rm Edd}$.  Hence, even
in Eddington units, Sgr~A* is considerably more underluminous than
M87, for which $L\approx10^{42}\ {\rm erg\,s^{-1}}\approx 10^{-6}L_{\rm Edd}$.  
Here we collect various parameter estimates relevant for the putative
jet launching region near Sgr~A*.

We will find that Sgr~A* and M87 have roughly similar photon energy
densities, the frequency of the spectral peak, and magnetic field
strengths.  This leads to comparable (to within a factor of few)
physical (in cm) gap thicknesses in both cases.  However, the mass of
Sgr~A* black hole is about $1500$ smaller than in M87. This means that
in SgrA*, the gap thickness is comparable to large-scale size of the
system, and this leads to the suppression of the cascade.

\subsubsection{Jet Velocity and Orientation}
In the absence of an obvious jet, there are no direct measurements of
a putative jet velocity.  In contrast, if the jet feature reported in
\citet{Li-Morr-Baga:13} is real, it is necessarily oriented at a large
angle relative to the line of sight.  Here we will assume a similar,
moderate Lorentz factor for the jet ($\Gamma\approx2.3$) and an inclination
angle of $90^\circ$, where necessary, consistent with the assumed
parameters in \citet{Li-Morr-Baga:13}.

\subsubsection{Magnetic Field Strength Estimates \& Synchrotron Cooling Times} \label{sec:SgrA:MFS}
The detection of polarized emission at mm-wavelengths, and thus the
measurement of the Faraday rotation measure of
$\approx6\times10^5~{\rm rad\,m^{-2}}$, produces an estimate
for the accretion rate of
$\dot{M}=2\times10^{-9}\,M_\odot\,\yr^{-1}{-}2\times10^{-7}\,M_\odot\,\yr^{-1}$, 
depending on the details of the accretion model assumed
\citep{2000ApJ...538L.121A,2007ApJ...654L..57M}.  
Following equations \eqref{eq:rhoMdot}--\eqref{eq:Brho} gives a
corresponding estimate of the possible confined jet magnetic field of
up to
\begin{equation}
B \approx \sqrt{\frac{2\beta c \dot{M}}{\Rg^2}}
\approx
10^{2-3}\,\G\,.
\end{equation}
In both cases, the accretion power, ranging from
$10^{38}{-}10^{40}\,\erg\,\s^{-1}$, exceeds the potential jet power by
two orders of magnitude, and imply the need for low radiative
efficiencies, i.e., $\eta_d\approx0.01{-}10^{-4}$.  These field
strengths are similar to those inferred from spectral fitting 
\citep[e.g.,][]{2003ApJ...598..301Y,Brod-Loeb:06a}.  For concreteness we will
assume the lower field strength, corresponding to the comparatively
higher radiative efficiency, i.e., we will assume 100~G on
the horizon and thus roughly 10~G near the stagnation surface.  As a
consequence, the synchrotron cooling timescale is similar to that of
M87:
\begin{equation}
t_{\rm sync}
\approx
4.0\times10^5 B_1^{-3/2} \nu_{11.5}^{-1/2}~\s\,.
\end{equation}
Since the gravitational radius of Sgr~A* is much smaller, so is the
apparent outflow timescale, comparable to
\begin{equation}
t_{\rm outflow}
\approx
2.4\times10^2 \left(\frac{r}{10\Rg}\right)~\s\,,
\end{equation}
and thus for Sgr~A* synchrotron cooling is completely irrelevant to
the evolution nonthermal particle population on horizon scales.

\subsubsection{Seed Photon Density Estimates}
\label{sec:seed-photon-density-sgr}
\begin{figure}
\begin{center}
\includegraphics[width=\columnwidth]{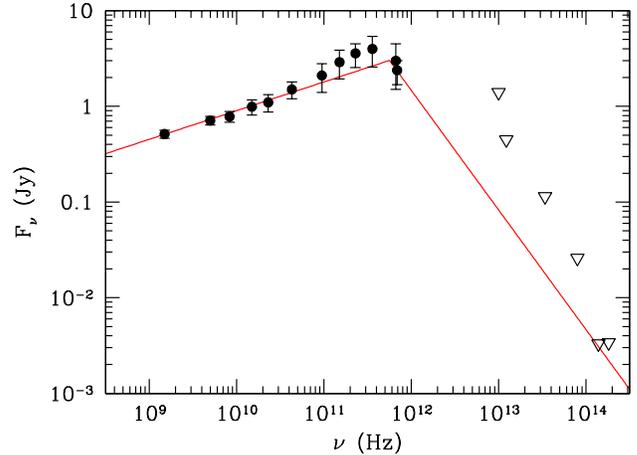}
\end{center}
\caption{Spectral energy distribution for Sagittarius A* from the
  radio to the optical, collected from 
  \citet[and references therein]{2004ApJ...606..894Y} and 
  \citet{2006PhDT........32M}.  Filled circles and open triangles correspond
  to flux measurements in which the source is resolved and unresolved,
  respectively; we take the filled circles as directly
  indicative of the near-horizon seed photon distribution and the open
  triangles as upper limits.  Errorbars indicate variability, not
  intrinsic measurement uncertainty.  For comparison, a simple broken
  power-law with high-frequency spectral index $\alpha=1.25$, as
  employed in the text, is shown for reference.}\label{fig:SGRASED}
\end{figure}
Unlike M87, Sgr~A* exhibits a clear sub-mm bump, often modeled
as a thermal disk component \citep{2003ApJ...598..301Y}, though also potentially
produced by the jet component associated with the jet launching region
\citep{2000A&A...362..113F,2002A&A...383..854Y,2013A&A...559L...3M}.
Nevertheless, above 1~mm the SED is well modeled by a power law with
$\alpha\approx1.25$ (see Figure \ref{fig:SGRASED}):  
\begin{equation}
\alpha\approx
\begin{cases}
0 & \nu<300~\GHz\\
1.25 & \nu\ge300~\GHz\,.
\end{cases}
\end{equation}
Furthermore, short-timescale
variability from the mm through the infrared imply that this emission
arises near the black hole, and thus may be identified with the soft
seed photons relevant for particle production and acceleration at the
putative jet stagnation surface.

The implied seed photon energy density above 1~mm is 
\begin{equation}
u_s\approx 1.5\times10^{-2}~\erg~\cm^{-3}\,,
\end{equation}
which is comparable to that implied in M87 (see Equation~\ref{eq:usm87}).
We will find, however, that the relevant seed photon population lies
above $10^{12}~\Hz$ (wavelengths shorter than $0.26~\mm$), for which
the energy density is reduced by roughly a factor of $5$.  As we will
see below, this leads to about a factor of $3$ increase in the physical thickness of
the gap (in cm) in comparison to M87. This turns out to lead a qualitative
change in the cascade operation: since Sgr~A* black hole mass is
$\sim1500$ times smaller than in M87, Sgr~A* gap thickness comes out to be
comparable to the size of black hole, and this suppresses the cascade.

\subsubsection{Implied Nonthermal Particle Properties}
\label{sec:impl-nonth-part}
The roughly power-law, declining SED of Sgr~A* in the sub-mm again
justifies assuming the emission is well characterized by optically
thin emission from a soft population of nonthermal electrons.  In this
case, $dn/d\gamma\propto\gamma^{-3.5}$, quite similar to M87.  With a
source size of $37~\muas$, equating the expression in Equation
(\ref{eq:Fmm}) to the observed millimeter flux of 3~Jy yields a
nonthermal particle density estimate of 
\begin{equation}
n\approx 10^6~\cm^{-3}.  
\label{eq:n_sgra}
\end{equation}
This is much
higher than for M87 as a direct result of the much smaller mass, and
therefore more compact emission region.  

Given the previous magnetic field estimate, the typical Lorentz
factors are expected to be of order $10^2$. While
the infrared emission requires considerably larger $\gamma$'s, it is
also highly variable and thus potentially associated with additional
dissipative events.

\subsection{Particle Acceleration at the Stagnation Surface}
As might be expected given the similarities in the soft photon
densities, the inverse-Compton limited Lorentz factor is about a
factor of $2$ smaller than that found in M87 (see Equation~\ref{eq:gammamax};
see Equation~\ref{eq:gammamaxm87} for M87 value):
\begin{equation}
\gamma_{\rm max}
\approx
7.6\times10^{8}\,.
\end{equation}
The length scale over which this is attained is (see
Equation~\ref{eq:ellIC})
\begin{equation}
\ell_{\rm IC} \approx 8.4\times10^{10}~\cm \approx 0.13\Rg\,.
\end{equation}
This value in cm is larger by about a factor of $2$ and in units of gravitational
radius by about a factor of $2300$ than the M87 value (see Equation~\ref{eq:ellicm87}).
As a consequence of the moderately lower $\gamma_{\rm max}$ the
threshold energy for seed photons which are able to pair produce on
the subsequently up-scattered inverse-Compton gamma is moderately higher,
\begin{equation}
\epsilon_{\rm th} = 2.7~\meV
\quad\Rightarrow\quad
\lambda_{\rm th}\approx0.46~\mm\,,
\end{equation}
which is pushing into the far-infrared.  Hence, unlike in M87, for Sgr
A* the soft photons responsible for pair production in the gap are
significantly above the spectral break.

Due to the larger $\epsilon_{\rm th}$, the corresponding mean free path
to pair production is comparatively large   (see Equation~\ref{eq:ellgg}),
\begin{equation}
\ell_{\gamma\gamma}
\approx
6.6\times10^{12}~\cm
\approx
10\Rg\,.
\end{equation}
This value exceeds that for M87 by a factor of $\simeq4$ in cm and in
units of $r_g$ by a factor of $5900$ (see Equation~\ref{eq:ellggm87}).
This implies that at most, the gap pair catastrophe is limited to a single
generation.  This suggests that the global jet dynamics, will play a
significant role in the determining the structure of the gap.
Ignoring this potential complication, the equilibrium gap thickness is of
order (see Equation~\ref{eq:gapthickness}; see Equation~\ref{eq:deltam87} for M87 value)
\begin{equation}
\Delta \approx 1.1\times10^{12}~\cm \approx 1.7\Rg\,,
\end{equation}
itself larger than a gravitational radius (though given the wide
disparity between $\ell_{\rm IC}$ and $\ell_{\gamma\gamma}$ this may be
highly variable).  The associated density at the top of the gap is
(see Equation~\ref{eq:ngap})
\begin{equation}
n_{\rm g} \approx 2.5\times10^{-3}~\cm^{-3}\,,
\label{eq:ngsgra*}
\end{equation}
about an order of magnitude smaller than for M87 (Equation~\ref{eq:ngm87}).

In summary, the gap densities for Sgr~A* (Equation~\ref{eq:ngsgra*}) and M87
(Equation~\ref{eq:ngm87}) are within an order of magnitude
of each other because of the similarity of seed photon densities,
magnetic field strengths, and the peak frequency of the spectrum.
The order of magnitude smaller density in Sgr~A* comes from a factor of
few smaller values of magnetic field strength and above threshold
photon density for Sgr~A*, falling well short of the $n_\infty\simeq
10^6~\cm^{-3}$ inferred from radio observations (see
Sec.~\ref{sec:impl-nonth-part}).

\subsection{Post-Gap Cascade} \label{sec:SgrAPGC} 

Since $\ell_{\gamma\gamma}$ is much larger than the system scale
height at the gap, i.e., $\ell_{\gamma\gamma}\gg r\approx10\Rg$, there
is effectively no post-gap cascade.  This is because for compact emission
regions the pair-production optical depth falls significantly at large
radii for two reasons.  First,  the seed photon density drops $\propto
r^{-2}$ for sufficiently large $r$.  Second, the direction of
propagation of the seed photons and gamma rays become increasingly
aligned as they travel far from their origin, reducing their
center-of-mass energy and increasing the energy threshold for pair
production $\propto r^2$.  Hence a gamma ray that fails to annihilate
in the first mean free path is unlikely to ever do so.
 
Thus,
$n_\infty\approx n_{\rm g}\approx2.5\times10^{-3}~\cm^{-3}$, implying
that in this case it is likely that Sgr~A* emission is likely
  coming from the accretion disk or from the disk outflow just outside
  of the highly magnetized jets
  \citep[e.g.,][]{2013A&A...559L...3M}. If a substantial fraction of
Sgr~A* arises from jet emission (but note that this does not
  have to be the case: in fact, SgrA* might not have any jet at all),
some other particle loading and acceleration mechanisms must be
occurring, such as gas entrainment from the accretion disk and the
acceleration of electrons via reconnection and/or shocks.

\subsection{Gamma-Ray Emission}
According to Equations~\eqref{eq:felep} and \eqref{eq:Llep}, the pair cascade luminosity scales
as $\propto M^2n_g\gamma_{\rm max,IC}$. Since black hole mass is
smaller by $3$ orders of magnitude, density is smaller by an order
of magnitude, and $\gamma_{\rm max,IC}$ is smaller by a factor of two,
the pair cascade luminosity is roughly $\sim7$ orders of magnitude
smaller than in M87 (Equation~\ref{eq:lcascadem87}), or
$\approx 3\times10^{35}\ {\rm erg\,s^{-1}}$.
Since $\gamma_\infty\approx10^7$ the overwhelming majority of this is deposited
in gamma rays beamed along the jet with typical energies of
$\gamma_{\rm max} m_e c^2 \approx 400~\TeV$.

Based on the overall gamma-ray luminosity it is tempting to relate
this to the power in the 1~GeV--100~GeV gamma-ray 10~\kpc-scale jet
features found by \citet{Su-Fink:12}. Such an identification, however,
requires at least two additional mechanisms.  First, the gamma rays
must be reprocessed down in energy by between three and five orders of
magnitude.  Second, the emission must be scattered in a distributed
fashion along the jet to produce the observed extended feature.
Accordingly, we leave the discussion of potential signatures of
gamma-ray emission from Sgr~A*'s putative jet for future work.

\subsection{Self-Consistent Jet Solutions} \label{sec:SgrAJet}
We finish our discussion of Sgr~A* with a consideration of potentially
self-consistent jet solutions.  While it has been found above that a jet
confined by an observationally motivated RIAF models is expected to
produce little intrinsic emission, this need not be the case if the
RIAF picture is relaxed.  

Increasing the magnetic field increases the efficiency of the
inverse-Compton pair catastrophe.  For $B\gtrsim200\,\G$
$\ell_{\gamma\gamma}$ becomes small in comparison to the gap height,
and thus multiple generations of pairs can be sustained resulting in a
well-defined inverse-Compton pair catastrophe.  
Simultaneously, the number of pairs required to reproduce the observed
sub-mm emission decreases.  For $B\approx10^4\,\G$ the two are
approximately equal, suggesting the existence of a self-consistent jet
model similar to that found for M87.  As with weaker fields
synchrotron cooling may continue to be ignored (this is true for
$B\lesssim1.5\times10^4\,\G$).

However, note that $10^4\,\G$ would be much larger than even optimistic
estimates for the magnetic field strength in Sgr A*.  The implied jet
power would be roughly $L_{\rm EM} \gtrsim 10^{41}\,\erg\,\s^{-1}$, five
orders of magnitude larger than Sgr A*'s bolometric luminosity.  While
such large luminosities have been implicated in Sgr A*'s past \citep[e.g.,
the Fermi bubbles][]{2010ApJ...724.1044S}, it is difficult to imagine
sufficiently low radiative efficiencies to hide such jet power 
in the present epoch.  

Perhaps more problematic is the need to confine such a large magnetic
flux.  This is ostensibly done via an accretion flow, which implies a
mass accretion rate of
\begin{equation}
\dot{M} \approx \frac{B^2 \Rg^2}{2\beta c} 
\gtrsim 10^{-5} \,M_\odot\,\yr^{-1}\,,
\end{equation}
comparable to the Bondi rate for Sgr~A*.  As discussed in Section
\ref{sec:SgrA:MFS}, such high accretion rates are already excluded by
polarimetric observations.  Moreover, the corresponding 
radiative efficiency of the accretion flow would need to be
exceedingly small: $\eta_d\lesssim10^{-6}$, much lower than the already
small efficiencies implicated in RIAF models.  

For these reasons we consider a self-consistent jet model driven by
a gap-powered inverse-Compton pair catastrophe to be highly disfavored
as an explanation of Sgr~A*'s current sub-mm emission.  Nevertheless,
this may not be the case for Sgr~A*'s recent past, where the
luminosity of Sgr A* has been inferred to be considerably higher.

\section{Discussion}
\label{sec:discussion}
\subsection{Source of seed photons}
Our model relies on the presence of seed photons with energies
$\epsilon_s \approx 1$~meV, or wavelengths $\lambda \approx 1$~mm, to
sustain the cascade. The spectra of Sgr~A* and M87 above the peak of
the SED give us an estimate of the seed photon density at the source.
However, so far we have been agnostic about the origin of these
photons: whereas in the case of Sgr~A* they could plausibly only come
from the accretion disk, in M87 they could come from the disk,
the jet, or both \citep[see, e.g.,][]{2009ApJ...697.1164B}.

In our model of M87, the synchrotron luminosity of the jet, which is
lit up by the leptons accelerated in the cascade, naturally explains
the SED of M87 above its peak at $\epsilon_s\gtrsim1$~meV. This is
because the pair cascade catastrophe naturally leads to pair density
of $n_\infty\approx15\ {\rm cm}^{-3}$ and Lorentz factor
$\gamma_\infty\approx10^4$ (see \S\ref{sec:post-gap-cascade} and
\S\ref{sec:total-power-cascade}): 
both of these values are comparable to the those required to account
for M87's radio emission above the peak of the SED
\citep{2009ApJ...697.1164B}.  This means that in principle no external
source of seed photons is required: the cascade is self-sustaining,
similar to polar cascades in pulsar magnetospheres (see, e.g.,
\citealt{2013MNRAS.429...20T}).  That is, given a stray seed photon
(e.g., from an accretion disk), the cascade fires up and produces
energetic electrons that then cool and produce sub-mm photons that can
serve as seed photons for the cascade.

\subsection{Comparison to other work}

An alternative process of pair production in jets is annihilation of
gamma-rays from the disk
\citep{2011ApJ...730..123L,2011ApJ...735....9M}.  In this scenario, a
sub-mm photon, emitted by the disk, undergoes a sequence inverse
Compton scattering events with disk's own hot electrons. These
interactions upscatter the photon to gamma-ray energies, and
eventually above pair production threshold.  Collisions of these
gamma-rays in the funnel mass-load the jet with pairs.

In order for this process to lead to an interestingly large number
density of pairs (larger than the Goldreich-Julian density, $n_{\rm
  GJ}$), the following two conditions must be met: (i) the mm-emission
of the disk must be compact and marginally optically thick to Thompson
scattering; (ii) disk electrons must be very hot, so that on average
no more than $\approx1.5$ Compton scatterings are needed to reach
pair-production threshold. To satisfy these requirements, disk
radiative efficiency is assumed to be high, $\eta_d\approx 0.3$
\citep{2011ApJ...735....9M}. 

In contrast, our model allows for standard interpretation of a
radiatively inefficient accretion flow in M87, $\eta_d\ll1$, which we believe is a
more natural assumption. The seed photons around the SED peak are
produced in the jet, outside of the disk proper, and thus they have a
lower probability of being up-scattered to gamma-ray energies by disk
electrons.  Both of these factors combined render ineffective the
process of jet pair loading due to $\gamma{-}\gamma$ collisions in the
context of our model.

\citet{2011ApJ...730..123L} considered pair cascade at the stagnation
surface of black hole magnetosphere and focused on the gamma-ray
emission from the resulting leptons. In our work we take a broader
approach and construct quantitative 1D numerical simulations of the
post-cascade evolution. We consider both the direct gamma-ray signal as well the
synchrotron emission of the leptons produced in the cascade. In fact,
we argue that the seed photons are produced in the cascade itself and
not the accretion flow. We use the observed spectrum of M87 to
determine the seed photon spectrum and eliminate the sensitivity of
the model to the uncertain details of the accretion flow.

\section{Conclusions}
\label{sec:conclusions}

Particle creation at the stagnation surface of Poynting-dominated jets
presents a natural mechanism for filling black hole jets with
nonthermal particles near the horizon, as required by recent mm-VLBI
observations of M87.  This model provides an excellent quantitative
description of a polar electron-positron cascade that can fill black
hole jets with relativistically-hot plasma with sufficient abundance
to explain the observed jet emission in M87.

Within the stagnation surface of astrophysical jets, large electric
fields are anticipated, arising due to the charge starvation within the
jet.  These are capable of accelerating stray leptons to extremely
high Lorentz factors, limited by inverse-Compton cooling on the
ambient soft-photon background, due to local synchrotron emission
within the jet or through their production in an accretion disk.  The
up-scattered gamma rays are sufficiently energetic to initiate a pair
catastrophe that ultimately self-consistently determines the structure
and content of the jet near the stagnation surface, i.e., within the
gap, beyond which sufficient charge densities exist to screen the
accelerating electric fields.  Typical gap thicknesses are much
smaller than the gravitational radius, and thus largely independent of
the global jet structure.

Subsequent to exiting the gap, the leptons typically have very large
Lorentz factors, and thus participate in a post-gap inverse-Compton
cascade.  While this does not increase the charge density, and
therefore does not affect the jet structure, it does significantly
increase the lepton density and decrease their specific energy.
Typical lepton density enhancements can reach $10^3$, and thus the
post-gap cascade can have a dramatic impact upon the particle content
of astrophysical jets.  The resulting pairs are then natural
candidates for the relativistic lepton populations that produce the
synchrotron emission in the cores of radio AGN.
In particular, particle creation at the stagnation surface provides an
excellent quantitative explanation of the horizon-scale emission
observed in M87.
In contrast, it vastly under produces the lepton content of Sgr~A*'s
emission region, consistent with the lack of an obvious jet in that
source.

Gamma-ray luminosity produced by the stagnation surface pair cascade
considerably exceeds that detected by \emph{Fermi}/LAT for M87, though
it is similar to those found for gamma-ray bright blazars.  However,
this naturally results from the strong beaming along the jet of the
up-scattered emission, implying a viewing inclination larger than
$\approx9^\circ$.  Thus, within the context of the stagnation surface
pair catastrophe model we present here, the lack of bright high-energy
gamma-ray emission from M87 favors the larger inclination implied by
high-frequency radio observations.  Nevertheless, future work should
compute the angular dependence of the emergent inverse-Compton spectra
and the potentially strong constraints on jet orientation that may
result.

\acknowledgments The authors thank Dimitrios Giannios, Ramesh Narayan
and Alexander Philippov for helpful discussions.  The authors would
also like to thank the anonymous referee for many helpful suggestions
that have resulted in a much improved presentation.
A.E.B.~receives financial support from Perimeter
Institute for Theoretical Physics and the Natural Sciences and
Engineering Research Council of Canada through a Discovery Grant.
A.T. was supported by a Princeton Center for Theoretical Science
Fellowship and by NASA through the Einstein Fellowship Program, grant
PF3-140131. The simulations presented in this article used
computational resources supported by XSEDE allocation TG-AST100040 on
NICS Kraken and Nautilus and TACC Lonestar, Longhorn, Ranch, and
Stampede.  A.T. thanks Perimeter Institute for hospitality and
financial support of three visits to Perimeter Institute during which most
of the work on this project was carried out.

\appendix

\section{Relativistic Counter-Streaming Pair Two-Stream Instability} \label{app:csts} 
One might be concerned with the potential for heating of the
counter-streaming electron and positron beams, ultimately radiated
away as synchrotron emission.  Here we estimate the growth rate of
this variant of the classic two-stream instability.  Note that in this
case we really do have a one-dimensional problem; particles are
strongly confined to field lines and thus can only participate in
collective motions in the direction of motion.

The electrons and positron motions are governed by the Boltzmann
equation, which upon some simplification gives the following relation
for the charge perturbations in terms of the unperturbed lepton
distribution functions (see, for example, the appendix of Broderick et
al. 2012):
\begin{equation}
\begin{aligned}
\rho_1 
&= 
\frac{i e^2}{m} \bmath{E}\cdot \int \left(f_0^++f_0^-\right)
\frac{\bmath{k}-\bmath{k}\cdot\bmath{v}\bmath{v}}{\gamma(\omega-\bmath{k}\cdot\bmath{v})^2}\,
d^3\!p\\
&= 
\frac{i e^2}{m} Ek \cdot \int
\frac{f_0^++f_0^-}{\gamma^3 (\omega-kv)^2}\,
d^3\!p\,,
\end{aligned}
\end{equation}
where in the second expression we have assumed that all the vectors
are co-linear due to the afore mentioned dimensionality of the
problem.  Inserting
$f_0^\pm = (n/2) \delta^3(\bmath{p}\pm\bmath{p_0})$
and performing the trivial integrals results in
\begin{equation}
\rho_1 = \frac{i e^2n}{2m\gamma^3} Ek\left[
\frac{1}{(\omega-kv)^2}
+
\frac{1}{(\omega+kv)^2}
\right]\,.
\end{equation}
From Gauss law ($\bmath{\nabla}\cdot\bmath{E}=4\pi\rho$) we then find
\begin{equation}
ikE = 4\pi\rho_1 = 
\frac{4\pi i e^2n}{m\gamma^3} \frac{Ek}{2}
\left[
\frac{1}{(\omega-kv)^2}
+
\frac{1}{(\omega+kv)^2}
\right]\,,
\end{equation}
or
\begin{equation}
1 - \frac{\omega_P^2}{2\gamma^3}
\left[
\frac{1}{(\omega-kv)^2}
+
\frac{1}{(\omega+kv)^2}
\right]
=
1 - \frac{\omega_P^2}{2\gamma^3}
\frac{\omega^2+(kv)^2}{\left[\omega^2-(kv)^2\right]^2}
=
0\,,
\end{equation}
or, with $\bar{k}=kv\gamma^{3/2}/\omega_P$ and
$\bar{\omega}=\omega\gamma^{3/2}/\omega_P$,
\begin{multline}
\bar{\omega}^4 - (2 \bar{k}^2+1)\bar{\omega}^2 
-\bar{k}^2(1-\bar{k}^2) = 0\\
\begin{aligned}
\Rightarrow\quad
\bar{\omega}^2
&=
\bar{k}^2+\frac{1}{2} \pm
\sqrt{\left(\bar{k}^2+\frac{1}{2}\right)^2
+
\bar{k}^2(1-\bar{k}^2)
}\\
&=
\bar{k}^2+\frac{1}{2} \pm
\sqrt{2\bar{k}^2+\frac{1}{4}}\,.
\end{aligned}
\end{multline}
This is minimized for the second root (the $-$ sign), when
\begin{equation}
1 - \frac{1}{\sqrt{2\bar{k}^2+1/4}}
= 0
\quad\Rightarrow\quad
\bar{k}^2 = \frac{3}{8}
\quad\Rightarrow\quad
\bar{\omega}^2 = -\frac{1}{8}\,.
\end{equation}
Thus, the growth rate of the relativistically counter-streaming pair
instability is,
\begin{equation}
\Gamma_{\rm CSI} = \Im(\omega)
=
\frac{\omega_P}{\sqrt{8\gamma^3}}\,.
\end{equation}

Within the gap the density given in Equation (\ref{eq:ngap}) sets a
characteristic upper limit,
\begin{equation*}
n_{\rm g} = \frac{\bmath{\nabla}\cdot\bmath{E}}{4\pi e} \approx \frac{E}{4\pi e \Delta}
=
3.3\, R_{15}^{3/2} \Omega_{F,-4}^{3/2} B_{2}^{3/2}
u_{s,0}^{1/2}
\eta_{\rm th}^{-1/2}
~\cm^{-3}\,.
\tag{\ref{eq:ngap}}
\end{equation*}
Note that here the characteristic length scale is $\Delta$, i.e., much
smaller than the distance to the stagnation surface, $10r_g$.
The corresponding plasma frequency is then
\begin{equation}
\omega_P\approx 1.0\times10^5
R_{15}^{3/4} \Omega_{F,-4}^{3/4} B_{2}^{3/4} u_{s,0}^{1/4} \eta_{\rm th}^{-1/4}
~\Hz\,,
\end{equation}
and thus setting $\gamma=\gamma_{\rm max}$,
\begin{equation}
\Gamma_{\rm CSI} \approx 
4.2\times10^{-9}
u_{s,0}\eta_{\rm th}^{-1/4}
~\Hz\,,
\end{equation}
which is typically exceedingly slow in comparison to the
light-crossing time of the gap.  The corresponding cooling length is
$\ell_{\rm CSI} \approx c/\Gamma_{\rm CSI}$.

\section{One-Dimensional, Approximate Gap Structure} 

The gap structure is set by the competition between pair
production and the screening of the externally applied 
electric field by the resulting charge separation.  In principle,
this requires a solution to the fully non-linear coupled
Maxwell-Boltzmann equations, including inverse Compton
scattering. Particle-in-cell (PIC) simulations could be one such
approach. However, in practice, a rough estimate of the gap
structure may be obtained via an approximate form of the
Maxwell-Boltzmann equations, in which the non-linearity is
manifested by boundary conditions.

Here we explicitly construct the solutions to the simple gap particle
population model listed in Section \ref{sec:gapdensity}.  
We assume that the electric field within the gap $E_g$ is sufficient to
instantaneously drive particles to their inverse Compton limited
velocities, and thus the particle distributions are essentially
monoenergetic.  The resulting particle distribution will evolve to
screen the externally applied $E_g$, preventing further acceleration
and saturating the growth of the gap lepton density.  Note that in
this way the linear model encapsulates the nonlinearity inherent in
the full Maxwell-Boltzmann equations.

For concreteness, we assume that the electric field is positive in the
upward direction, corresponding to the direction of the guiding
magnetic field line, in which case the electrons and positrons accelerate
downward and upward, respectively.  Then, the number densities of the
pairs ($n_{e^+}$, $n_{e^-}$) and upward and downward propagating
gamma-rays ($n_{\gamma^+}$, $n_{\gamma^-}$) evolve due to pair
production and inverse Compton scattering via:
\begin{equation}
\begin{aligned}
\dot{n}_{e^+} + c \partial_z n_{e^+}
&=
\alpha_{\gamma\gamma} c \left( n_{\gamma^+} + n_{\gamma^-} \right)\\
\dot{n}_{e^-} - c \partial_z n_{e^-}
&=
\alpha_{\gamma\gamma} c \left( n_{\gamma^+} + n_{\gamma^-} \right)\\
\dot{n}_{\gamma^+} + c \partial_z n_{\gamma^+}
&=
\alpha_{\rm IC} c n_{e^+}
-
\alpha_{\gamma\gamma} c n_{\gamma^+}\\
\dot{n}_{\gamma^-} - c \partial_z n_{\gamma^-}
&=
\alpha_{\rm IC} c n_{e^-}
-
\alpha_{\gamma\gamma} c n_{\gamma^-}\,.
\end{aligned}
\label{eq:gap}
\end{equation}
where $\alpha_{\gamma\gamma}\equiv\sigma_{\gamma\gamma} n_s$,
$\alpha_{\rm IC}\equiv\sigma_T n_s$, and $n_s$ is the number density of seed
photons.  Pairs are generated by both upward and downward propagating
photons, though as assumed travel only in their prescribed
directions.  Thus, the upward (downward) propagating gamma-rays are
produced by inverse Compton scattering by only the positrons
(electrons).

While Equation (\ref{eq:gap}) is manifestly one-dimensional, the
stagnation surface itself extends transversely over a macroscopic
scales, i.e., typically much longer than $\ell_{IC}$ or
$\ell_{\gamma\gamma}$.  The strong beaming of the up-scattered
gamma rays by the highly relativistic, accelerated leptons, implies
that regions that are separated by a transverse distance of more
$\Delta/\gamma$, where $\Delta$ is the gap height to be determined
below, execute independent pair catastrophes.  As a result, the
stagnation surface is comprised of a large number 
($\approx \gamma^2 R^2/\Delta^2$) of essentially independently
evolving one-dimensional, pair cascades, each described
by Equation (\ref{eq:gap}).

The pair cascade is inherently unstable, as will be shown in detail
in Section \ref{app:1dgapdyn} though is clear from the nature of the
quenching: pairs form and accelerate until the resulting charge
segregation screens the externally applied electric field.  As a
consequence, the microscopic gap structure, i.e., that within a single
causally connected tube, is highly variable.  However, the macroscopic
gap structure, comprised by very many causally connected tubes, is
well described by an ensemble average over microscopic tubes at
different stages of the discharge cycle, i.e., the time-averaged
structure of a single microscopic region.  Due to the linearity of
Equation (\ref{eq:gap}), this corresponds immediately to the
stationary solution, which we address first.

\subsection{Stationary Structure}\label{app:1dgap}
Assuming stationarity, all the time derivatives vanish, and Equation
(\ref{eq:gap}) is a set of coupled ordinary differential equations for
the gap structure: 
\begin{equation}
\begin{aligned}
\partial_z n_{e^+}
&=
\alpha_{\gamma\gamma} \left( n_{\gamma^+} + n_{\gamma^-} \right)\\
- \partial_z n_{e^-}
&=
\alpha_{\gamma\gamma} \left( n_{\gamma^+} + n_{\gamma^-} \right)\\
\partial_z n_{\gamma^+}
&=
\alpha_{\rm IC} n_{e^+}
-
\alpha_{\gamma\gamma} n_{\gamma^+}\\
- \partial_z n_{\gamma^-}
&=
\alpha_{\rm IC} n_{e^-}
-
\alpha_{\gamma\gamma} n_{\gamma^-}\,.
\end{aligned}
\label{eq:gapstatic}
\end{equation}
Choosing $z=0$ to lie in the gap center, symmetry requires that 
\begin{equation}
n_{e^+}(z) = n_{e^-}(-z) \equiv f(z)
\label{eq:nep}
\end{equation}
and similarly
\begin{equation}
n_{\gamma^+}(z)=n_{\gamma^-}(-z) \equiv g(z).
\label{eq:nem}
\end{equation}
Thus, Equations~\eqref{eq:gapstatic} reduce to
\begin{align}
f' &= \alpha_{\gamma\gamma} \left[ g(z) + g(-z) \right]\\
\intertext{and}
g' &= \alpha_{\rm IC} f - \alpha_{\gamma\gamma} g\,,
\end{align}
where for compactness we have denoted $\partial_z$ with primes.
Taking the second derivative of the second equation and inserting the
first, we obtain the following for $g$:
\begin{equation}
g'' + \alpha_{\gamma\gamma} g' - \alpha_{\rm IC} \alpha_{\gamma\gamma}\left[ g(z) + g(-z) \right] = 0\,.
\end{equation}
Noting that $g'(-z)=-[g(-z)]'$ and $g''(-z)=[g(-z)]''$, this implies
\begin{equation}
\begin{gathered}
\left[g(z)+g(-z)\right]'' - 2\alpha_{\rm IC} \alpha_{\gamma\gamma}\left[ g(z) + g(-z) \right] = 0\\
\alpha_{\gamma\gamma}\left[g(z)-g(-z)\right]' = 0\,,
\end{gathered}
\end{equation}
from which, with the required symmetry, we obtain the solutions
\begin{equation}
g(z) = A \cosh\left(\sqrt{2\alpha_{\rm IC}\alpha_{\gamma\gamma}}\, z\right)\,.
\end{equation}
We may now construct $f$ by integration:
\begin{equation}
f = A \sqrt{\frac{\alpha_{\gamma\gamma}}{2\alpha_{\rm IC}}}
\sinh\left(\sqrt{2\alpha_{\rm IC}\alpha_{\gamma\gamma}}\, z\right) + B\,.
\end{equation}
The contribution to the electric field from the pairs within the gap
may be obtained directly by integrating Gauss' law,
\begin{equation}
\begin{aligned}
E'_\parallel &= 4\pi e \left(n_{e^+}-n_{e^-}\right) = 4\pi e \left[f(z) - f(-z)\right]\\
\Rightarrow\quad
E_\parallel &= 
\frac{4\pi e A}{\alpha_{\rm IC}}
\cosh\left(\sqrt{2\alpha_{\rm IC}\alpha_{\gamma\gamma}}\,z\right)
+
C\,.
\end{aligned}
\end{equation}
When $A=0$, i.e., the charge density vanishes, the total electric
field is simply that due to the gap, $E_g$.  The particle cascade
saturates when the gap is closed, i.e., $E_\parallel\approx0$ at the center, and
$E_\parallel=E_g$ at the gap boundary, where it is presumably screened by MHD
processes, taken to lie at a height $z_g$.  The first condition gives 
\begin{equation}
C = - \frac{4\pi e A}{\alpha_{\rm IC}}\,.
\end{equation}
The second then sets $A$ in terms of the gap size:
\begin{equation}
A = \frac{\alpha_{\rm IC} E_g }{4\pi e
\left(\cosh\delta-1\right)}\,,
\end{equation}
where 
$\delta\equiv\sqrt{2\alpha_{\rm IC}\alpha_{\gamma\gamma}} z_g=2z_g/\Delta$,
where $ \Delta\equiv\sqrt{2\ell_{\rm IC}\ell_{\gamma\gamma}}$ is the
characteristic gap length scale. 

Because all particle production is assumed to occur within the gap,
only outgoing particles exist above and below the gap boundaries.
The condition that $f\ge0$ everywhere within the gap then requires
that the density of each species vanishes at a corresponding
boundary, set by their respective directions of motion, i.e., the
positron and electron densities vanishes at the bottom and top of the
gap, respectively.  This implies,
\begin{equation}
B=A \sqrt{\frac{\alpha_{\gamma\gamma}}{2\alpha_{\rm IC}}}
\sinh\delta\,,
\end{equation}
from which we infer the particle densities at the top of the gap,
\begin{equation}
n_{e^+}(z_g)
\approx \frac{\sqrt{2\alpha_{\rm IC}\alpha_{\gamma\gamma}}\, E_g}{4\pi e}
\frac{\sinh\delta}
{\cosh\delta-1}\,.
\label{eq:ne}
\end{equation}

It is clear from the above that the natural gap scale is
$\Delta$.  However, the density at the top of the 
gap is a weak function of $\delta$, with
$\sinh\delta/(\cosh\delta-1)$ dropping to $2$ at $\delta\simeq1$ and
falling below $1.3$ for $\delta>2$.  Thus, independent of the precise
gap thickness, the density at the top of the gap is
\begin{equation}
n_{e^+} 
\approx 2 \frac{E_g}{4\pi e \Delta}\,,
\label{eq:ngapappendix}
\end{equation}
which is directly comparable to Equation (\ref{eq:ngap}).
Kinetic plasma simulations are desirable to verify this
linearized model but are beyond the scope of this work.

Equation~\eqref{eq:ngapappendix} gives a value of charge density that
is a factor $R/\Delta = 10r_g/\Delta\sim10^5$ times greater than the
Goldreich-Julian density, $\rho_{\rm GJ}\sim \Omega B/2\pi c$.
Consequently, the gap structure implies a large pair multiplicity that
is, however, not unusual for magnetospheric cascades, e.g., in the 
context of pulsars \citepalias{2013MNRAS.429...20T}. In fact, the
multiplicity $\eta = R/\Delta$ implied by
Equation~\eqref{eq:ngapappendix} is in good agreement with the simulation
results of \citetalias{2013MNRAS.429...20T} (see the discussion
after Equation~\ref{eq:ngap}).

\subsection{Gap Dynamics}\label{app:1dgapdyn}
The gap model described above is already linearized, simplifying the
discussion of its dynamics considerably.  Here we look at small,
harmonic perturbations about the steady state configuration.
We have some freedom to measure time and distance in convenient units,
which to simplify the following expressions, we do in
$\ell_{\gamma\gamma}=1/\alpha_{\gamma\gamma}c$.  Then, defining
$\varpi\equiv\omega/\alpha_{\gamma\gamma}c$,
$\kappa\equiv k/\alpha_{\gamma\gamma} c$, and 
$\zeta\equiv\alpha_{\rm IC}/\alpha_{\gamma\gamma}$, the equation for
the perturbations becomes
\begin{equation}
\begin{aligned}
-i(\varpi - \kappa) \delta n_{e^+}
&=
\delta n_{\gamma^+} + \delta n_{\gamma^-} \\
-i(\varpi + \kappa) \delta n_{e^+}
&=
\delta n_{\gamma^+} + \delta n_{\gamma^-} \\
-i(\varpi - \kappa) \delta n_{\gamma^+}
&=
\zeta \delta n_{e^+}
-
\delta n_{\gamma^+}\\
-i(\varpi + \kappa) \delta n_{\gamma^-}
&=
\zeta \delta n_{e^-}
-
\delta n_{\gamma^-}\,.
\end{aligned}
\label{eq:gapdyn}
\end{equation}
This may be trivially rearranged to define the modes as a standard
eigenmode problem:
\begin{equation}
\left(
\begin{matrix}
\kappa & 0 & i & i\\
0      & -\kappa & i & i\\
i\zeta   & 0 & \kappa-i & 0\\
0      & i\zeta & 0 & -\kappa-i
\end{matrix}
\right)
\left(
\begin{matrix}
\delta n_{e^+}\\
\delta n_{e^-}\\
\delta n_{\gamma^+}\\
\delta n_{\gamma^-}
\end{matrix}
\right)
=
\varpi
\left(
\begin{matrix}
\delta n_{e^+}\\
\delta n_{e^-}\\
\delta n_{\gamma^+}\\
\delta n_{\gamma^-}
\end{matrix}
\right)\,.
\end{equation}
The associated four mode-specific dispersion relations are
\begin{equation}
\varpi = -\frac{i}{2}\left[ 1 \pm \sqrt{1+4\zeta-4\kappa^2\pm4\sqrt{\zeta^2-(1+4\zeta)\kappa^2}}\right]\,,
\end{equation}
where every combination of the two $\pm$ may be chosen.  Two of the
four modes are unstable ($--$ and $-+$), both of which correspond to
the pair catastrophe (related to the symmetric and anti-symmetric
components).  The maximum growth rate is
\begin{equation}
\left.\Gamma_{--}\right|_{\kappa=0} \equiv \alpha_{\gamma\gamma} c \Im(\varpi_{--})
=
\frac{\sqrt{\alpha_{\gamma\gamma}^2+8\alpha_{\gamma\gamma}\alpha_{\rm IC}}-\alpha_{\gamma\gamma}}{2}c\,.
\end{equation}
Thus, the growth time, $\Gamma_{--}^{-1}$, is roughly the light
crossing time of the open gap.  The physical origin of the instability
is easy to understand: additional leptons induce additional
inverse Compton scattered photons which, in turn, produces additional
electrons, causing the process to exponentiate.

\section{Direct Inverse Compton Flux Density Estimates} \label{sec:DICFDE}
Here we construct a simple model with which to estimate the direct
inverse Compton signal from the gap-accelerated leptons.  The critical
input assumption is that the leptons evolve solely due to
inverse Compton cooling after leaving the gap.  Simplifying
assumptions, that certainly hold in the case of M87 and are likely to
hold generically, include a broken power-law distribution for the seed
photons and a uniform seed photon density over the cooling length
scales of interest.  As in section \ref{sec:NAE}, for compactness we
will measure energies in units of electron rest mass.

For concreteness, the seed photon distribution is take to be of the
form given in Equation (\ref{eq:spd})
\begin{equation}
\frac{dn_s}{d\epsilon_s}
=
\frac{u_s}{\epsilon_b^2} \frac{(\alpha-1)(1-\beta)}{\alpha-\beta}
\begin{cases}
(\epsilon_s/\epsilon_b)^{-(\beta+1)} & \epsilon_s\le\epsilon_b\\
(\epsilon_s/\epsilon_b)^{-(\alpha+1)} & \epsilon_s>\epsilon_b\,,
\end{cases}
\end{equation}
i.e., a broken power-law with a spectral indexes of $\alpha$ and
$\beta$ above and below the break, respectively.  The above is
normalized such that the seed photon energy density is $u_s$.

The post-gap lepton population has a height-dependent, mono-energetic
distribution:
\begin{equation}
\frac{dn}{d\gamma} = n_\infty \delta[\gamma-\gamma(z)]\,,
\end{equation}
where $\gamma(z)$ is set by inverse Compton cooling, and determined by
\begin{equation}
\frac{d\gamma}{dz} = - \frac{4\sigma_T u_s}{3 m c^2} \gamma^2\,,
\end{equation}
subject to the initial condition that at the top of the gap,
$\gamma=\gamma_\infty$.  While this may be trivially integrated, only
the above expression is required below.

Given a mono-energetic electron distribution, the energies of the
up-scattered gamma ray and the initial seed photon are related by
\begin{equation}
\epsilon_s = \frac{\epsilon}{2\gamma(\gamma-\epsilon)}
\quad\Rightarrow\quad
\frac{d\epsilon_s}{d\epsilon}
=
\frac{1}{2(\gamma-\epsilon)^2}\,,
\end{equation}
where we have expressed these in a slightly different form than usual,
obtaining the seed photon energy in terms of the gamma-ray energy.

The resulting gamma-ray flux is then
\begin{equation}
F_\epsilon
=
\int d\gamma dz\,
\sigma_T c \frac{dn}{d\gamma} 
\frac{dn_s}{d\epsilon_s} \frac{d\epsilon_s}{d\epsilon}
\epsilon \frac{r^2}{D^2}\,,
\end{equation}
where the factor of $d\epsilon_s/d\epsilon$ simply converts from per
unit seed photon energy to per unit gamma-ray energy.  The beaming
corrections (enhanced emission along the direction of the electron
momentum and reduced viewing area) cancel as described in the main
text.  Affecting the integral over $\gamma$ is trivial due to the
mono-energetic nature of the electron distribution, hence,
\begin{equation}
\begin{aligned}
F_\epsilon
&=
\sigma_T c n_\infty
\epsilon 
\frac{r^2}{D^2}
\int_{z_{\rm gap}}^\infty dz\,
\frac{dn_s}{d\epsilon_s} \frac{1}{2[\gamma(z)-\epsilon]^2}\\
&=
\sigma_T c n_\infty
\epsilon 
\frac{r^2}{D^2}
\int^1_{\gamma_\infty} d\gamma
\frac{dn_s/d\epsilon_s}{d\gamma/dz} \frac{1}{2(\gamma-\epsilon)^2}\\
&=
\frac{3 m c^3 n_\infty}{\epsilon_b}
\frac{(\alpha-1)(1-\beta)}{\alpha-\beta}
\frac{r^2}{D^2}\\
&\quad
\times
\left[
\left(\frac{\epsilon}{2\epsilon_b}\right)^{-\beta}
\int_{\gamma_b}^{\gamma_\infty} d\gamma
\left[\gamma\left(\gamma-\epsilon\right)\right]^{\beta-1}
\right.\\
&\qquad
\left.
+
\left(\frac{\epsilon}{2\epsilon_b}\right)^{-\alpha}
\int_1^{\gamma_b} d\gamma
\left[\gamma\left(\gamma-\epsilon\right)\right]^{\alpha-1}
\right]\,,
\end{aligned}
\end{equation}
where
\begin{equation}
\gamma_b 
= 
\frac{\epsilon}{2}
\left(1 + \sqrt{1 + \frac{2}{\epsilon \epsilon_b}}\right)
\approx
\sqrt{\frac{\epsilon}{2\epsilon_b}}\,,
\end{equation}
is the electron Lorentz factor that up scatters seed photons at the
break to the desired observed energy.  For the cases of interest, 
$\epsilon\ll 1/\epsilon_b$, and the term in the radical dominates as
shown.  In the limit that $\epsilon\ll\gamma_b$, the 
remaining integrals may then be approximated by
\begin{equation}
\begin{aligned}
\int_{\sqrt{\epsilon/2\epsilon_b}}^{\gamma_\infty} d\gamma
\left[\gamma\left(\gamma-\epsilon\right)\right]^{\beta-1}
&\approx
\left.\frac{\gamma^{2\beta-1}}{2\beta-1}\right|_{\sqrt{\epsilon/2\epsilon_b}}^{\gamma_\infty}\\
&\approx
\frac{1}{1-2\beta}
\left(\frac{\epsilon}{2\epsilon_b}\right)^{\beta-1/2}
\,,
\end{aligned}
\end{equation}
where we have assumed that $\beta<1/2$, and
\begin{equation}
\int_\epsilon^{\sqrt{\epsilon/2\epsilon_b}} d\gamma
\left[\gamma\left(\gamma-\epsilon\right)\right]^{\alpha-1}\\
\approx
\frac{1}{2\alpha-1}
\left(\frac{\epsilon}{2\epsilon_b}\right)^{\alpha-1/2}\,.
\end{equation}
Thus,
\begin{equation}
F_\epsilon
=
\frac{6 m^2 c^5 n_\infty}{\epsilon_b}
\frac{(\alpha-1)(1-\beta)}{(2\alpha-1)(1-2\beta)}
\frac{r^2}{D^2}
\left(\frac{\epsilon}{2\epsilon_b}\right)^{-1/2}
\end{equation}
where in the final expression the factors of $mc^2$ have been
reintroduced.  The typical flux at energy $\epsilon$ is then
\begin{equation}
\begin{aligned}
\epsilon F_{\epsilon}
&=
\int_\epsilon^\infty d\epsilon F_\epsilon\\
&\approx
12 m^2 c^5 n_\infty
\frac{(\alpha-1)(1-\beta)}{(2\alpha-1)(1-2\beta)}
\frac{r^2}{D^2}
\left(\frac{\epsilon}{2\epsilon_b}\right)^{1/2}\\
&\approx
12 \gamma_b m c^3 n_\infty
\frac{(\alpha-1)(1-\beta)}{(2\alpha-1)(1-2\beta)}
\frac{r^2}{D^2}
\,.
\end{aligned}
\end{equation}

\bibliography{jss.bib}
\bibliographystyle{apj}

\end{document}